\documentclass[twocolumn]{aastex702}
\usepackage{amsmath}
\usepackage{multirow}
\usepackage{rotating}

\usepackage{hyperref}
\usepackage{booktabs}
\usepackage{multirow}
\usepackage{hyperref}
\usepackage[capitalize]{cleveref}
\begin{document}
\title{\textit{Swift} gives a new \texttt{BAT-GLIMPSE}: Gamma-ray Localization using Imaging and Mosaic techniques for Pointing and Slew Epochs}

\author[0000-0003-0020-687X]{Samuele Ronchini}
\email{samuele.ronchini@gssi.it}
\correspondingauthor{samuele.ronchini@gssi.it}
\affiliation{Gran Sasso Science Institute (GSSI), I-67100 L'Aquila, Italy}
\affiliation{INFN, Laboratori Nazionali del Gran Sasso, I-67100 Assergi, Italy}
\author[0000-0002-4299-2517]{Tyler Parsotan}
\affiliation{Astrophysics Science Division, NASA Goddard Space Flight Center,Greenbelt, MD 20771, USA}
\email{a.b@example.com}
\author[0000-0001-5229-1995]{James DeLaunay}
\affiliation{Department of Astronomy and Astrophysics, The Pennsylvania State University, 525 Davey Lab, University Park, PA 16802, USA}
\affiliation{Center for Multimessenger Astrophysics, Institute for Gravitation and the Cosmos, Pennsylvania State University, University Park, PA 16802, USA}
\email{a.b@example.com}
\author[0000-0002-6745-4790]{Jamie A. Kennea}
\affiliation{Astrophysics \& Space Center, Schmidt Sciences, New York, NY 10011, USA}
\email{a.b@example.com}

\begin{abstract}
The Burst Alert Telescope (BAT) onboard the Neil Gehrels \textit{Swift} Observatory is capable of localizing $\gamma-$ray transients with arcminute precision, enabling rapid multi-wavelength follow-up. However, onboard triggering is disabled during spacecraft slews, preventing the autonomous detection and localization of transients occurring during these intervals.
We present here \texttt{BAT-GLIMPSE}, a fully autonomous, open-source pipeline for the low-latency localization of transient $\gamma-$ray sources in \textit{Swift}-BAT data. Making use of the \texttt{BatAnalysis} package, the pipeline combines coded-mask imaging and mosaic techniques, automatically selecting the appropriate analysis according to the spacecraft attitude and enabling localization searches during both pointing observations and spacecraft slews.
We validate the performance of \texttt{BAT-GLIMPSE} on a sample of 66 GRBs reported in GUANO circulars. The pipeline successfully recovers arcminute positions for 43 events, consistent with published localizations, with typical offsets of $\lesssim5$ arcminutes. Approximately $88\%$ of the GRBs occurring during spacecraft slews are recovered through imaging or mosaic analyses. During the fourth LIGO-Virgo-KAGRA observing run, the role of \texttt{BAT-GLIMPSE} was crucial in the search for $\gamma-$ray counterparts of gravitational waves, particularly in response to pre-merger alerts which triggered the slew of the \textit{Swift} spacecraft with extremely low latency. Operating synergistically with \texttt{NITRATES}, \texttt{BAT-GLIMPSE} fills the critical gap left by slew intervals and, together, the two pipelines are estimated to double the onboard arcminute-localization rate of \textit{Swift}-BAT, unlocking the full potential of the \textit{Swift} mission for time-domain and multi-messenger astrophysics.
\end{abstract}


\section{Introduction}
\label{sec:intro}

The Neil Gehrels \textit{Swift} Observatory, launched in 2004 \citep{gehrels2004swift}, has three on-board telescopes: the Burst Alert Telescope (BAT; \citealt{barthelmy_BAT}), the X-ray Telescope (XRT; \citealt{burrows_XRT}), and the Ultraviolet-Optical Telescope (UVOT; \citealt{roming_UVOT}). These three telescopes combined with the observatory's quick slewing capabilities have allowed \textit{Swift} to play a pivotal role in the study of a wide variety of astrophysical sources. At the start of the mission, \textit{Swift} spent an extended period of time pointed at various objects and conducting follow-up observations of $\gamma-$ray burst (GRB) triggers. As time has passed, the observatory has shifted to conducting more Target of Opportunity (ToOs) observations. These ToOs encompass any target the community identifies as scientifically compelling, including the time criticality of the observations. While this has contributed to \textit{Swift} being an indispensable observatory, it has come with downsides related to the reduced time spent in pointed observations of a given point in the sky and for the serendipitous discovery of high-energy transients.

The BAT is the primary driver of transient science on the observatory. BAT is a wide field of view (FOV) hard X-ray coded mask instrument operating in the 15-150 keV energy range. The coded mask technique allows it to trigger on transient events and localize them to $\sim 3$ arcminute regions on the sky, which is a capability few gamma-ray instruments have \citep{burns2023gamma}. This capability is critical for coordinating follow-up of these transients to better characterize them. In order to increase the number of localizations that BAT can provide, \cite{tohuvavohu2_guano} developed the GUANO ground system which allows for commanded downlinks of BAT time tagged event (TTE) data -- the data type that is typically produced for an on-board triggered event. The GUANO system allows BAT to save TTE data for external triggers from other instruments, allowing its use in the localization of these events. The \texttt{NITRATES} low latency search pipeline uses this dataset to perform a maximum-likelihood analysis and to quickly localize triggers even if the source was not in the FOV of BAT \citep{delaunay_nitrates}.

The \texttt{NITRATES} pipeline performs a search for a GRB-like signal around the trigger time of the external event over a range of time-scales (0.128 s to 16.384 s) and over several spectral model templates (ranging from soft to hard GRB-like spectra). The significance of a candidate signal is determined by the resulting test-statistic, $TS$, which is 2 times the difference between the log likelihood (LLH) of a background plus GRB signal model maximized with respect to the GRB parameters and the LLH of a background-only model. The $TS$ is often presented in terms of $\sqrt{TS}$, as the square root of this test-statistic is distributed similarly to a Gaussian signal to noise ratio (SNR) for background-only samples and makes it easier to compare to standard imaging results, which uses a Gaussian SNR to determine significance.
With respect to standard imaging, the enhanced sensitivity of \texttt{NITRATES} gives on average an increase in detectable volume equal to a factor $\sim4.3$. In the context of multi-messenger astrophysics, the interplay between GUANO and \texttt{NITRATES} has yielded significant results in the context of multi-messenger astrophysics, allowing for the first detection of a GRB afterglow in the mm band \citep{2022ApJ...935L..11L}, and for placing constraints on the electromagnetic emission from GW events \citep{2024ApJ...970L..20R,2025ApJ...980..207R,2026ApJ...998..171Z}. However, although \texttt{NITRATES} has significantly increased the number of transients localized by BAT, this capability is still limited to only time periods when the spacecraft is not slewing.

Slewing is a time period that limits the time in which BAT can trigger and localize sources, since the triggering capability in the onboard software is disabled whenever the spacecraft is slewing. This is meant to prevent spurious triggers due to bright sources moving through the BAT's FOV. Due to the increased number of ToOs that \textit{Swift} executes, the spacecraft spends an increased amount of time slewing which then reduces the amount of time in which BAT can potentially trigger on a transient, even if it is in the FOV of the instrument. Fortunately, the GUANO pipeline allows us to downlink TTE data during these time periods. With these data in hand, the transient can still be localized on the ground, as was first outlined by \cite{slew_survey_copete}.  

The \texttt{BatAnalysis} package \citep{2025ApJ...988...23P} is a new open source python package which allows for the flexible processing of BAT TTE data including image generation and the construction of mosaics. \texttt{BatAnalysis} is built using the standard HEASoft tools and allows users to generate images by performing a standard cross correlation of the detector plane image with the coded mask \citep{2025ApJ...988...23P}. These images are then projected onto a healpix map and are summed together to produce the mosaiced images with increased sensitivity to faint sources \citep{2025ApJ...988...23P, parsotan2023batanalysis}. We have created a pipeline that leverages both the \texttt{NITRATES} and \texttt{BatAnalysis} codes to conduct fast localizations of transients in the imaging domain, regardless of whether BAT is slewing during the time period of interest or not. 

In Section \ref{methods} we outline the implementation of the pipeline. Then in Section \ref{grb_sample} we outline the selection of GRBs that were identified as validation tests for the pipeline, though the pipeline was successfully used in a number of real time applications. In Section \ref{results} we show the results of running the pipeline on these archival GRBs and make comparisons to the more sensitive \texttt{NITRATES} forward folding pipeline. In Section \ref{sensitivity}, we show the sensitivity gain of the \texttt{BAT-GLIMPSE} and  \texttt{NITRATES} pipelines working in sync with one another. In Section \ref{ultra_swift} we describe the importance of the pipeline in the context of the \textit{ULTRA-Swift} project (Baylor et al., in prep.).
Finally, in Section \ref{conclusions} we discuss the implications of the results of the pipeline and future use.

\section{\texttt{BAT-GLIMPSE} implementation} \label{methods}

In this section we describe the analysis performed by \texttt{BAT-GLIMPSE} \citep{batglimpse}, from the trigger received by an external instrument up to the dissemination of the results. The repository is publicly available on GitHub\footnote{\url{https://github.com/samueleronchini/BAT-GLIMPSE}}. Most of the functions implemented here rely on modules of the \texttt{BatAnalysis} package \citep{2025ApJ...988...23P}. \texttt{BAT-GLIMPSE} runs the analysis on all the multi-messenger transients that trigger GUANO, including GRBs detected by other telescopes that failed to trigger BAT onboard, gravitational waves, X-ray transients localized by Einstein Probe, fast radio bursts and IceCube high-energy neutrinos. From the moment GUANO sends the request for data downlink, $\sim$ few hours are necessary to have the full recovery of data after one or more communications with ground stations. Once the data arrive at the \textit{Swift} data center, the analysis can start.

The event file is cleaned by removing glitches and bad time intervals. A detector mask file and an attitude file are created as well. If the attitude file does not cover the external trigger time, two attitude files are merged using the closest \textit{Swift} observation ID in time. The reader can refer to Sec. 7.1 of \cite{delaunay_nitrates} for further details about data cleaning.

\subsection{Search of time seeds}
\begin{figure}
    \centering
    \includegraphics[width=1\linewidth]{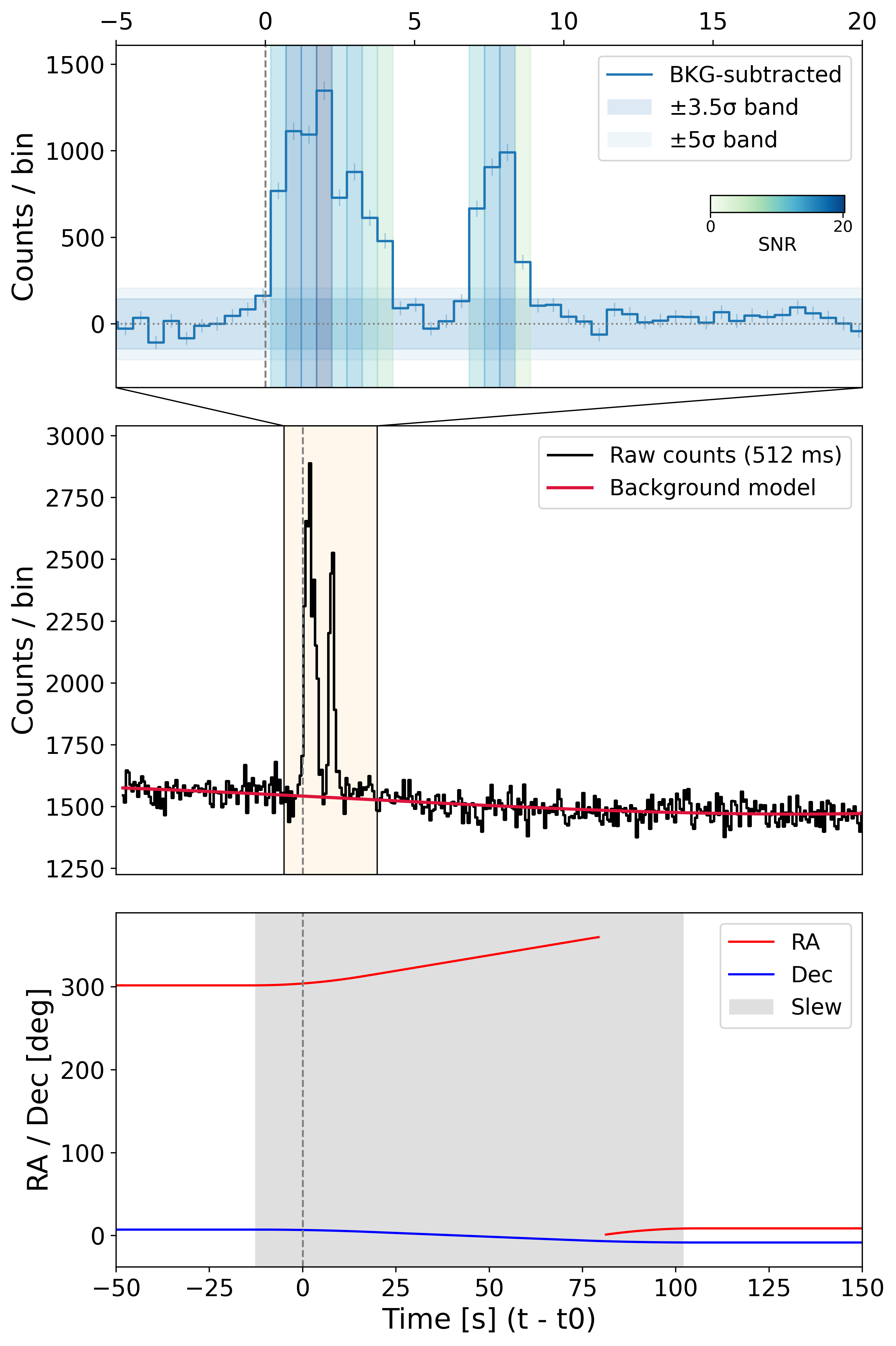}

    \caption{Example of light curve and background fitting during slew by \texttt{BAT-GLIMPSE}. Middle panel: raw light curve with background fit. The light yellow band indicates the time interval excluded from the fit, by default $[t_0-5\,s,t_0+20\,s]$. The red curve is the best fit background model. Top panel: inset with the background-subtracted light curve, with horizontal bands indicating 3.5$\sigma$ and $5\sigma$ uncertainty on the background level, and vertical bands highlighting the time bins where the SNR$>3.5$, colored by the value of SNR.  Bottom panel: time evolution of the \textit{Swift} boresight coordinates, where the slew is highlighted in gray. All panels are relative to GRB 221206B (GCN 33027, \citealt{2022GCN.33027....1D}).}
    \label{fig:lc}
\end{figure}
The pipeline first converts the filtered event file into a binned light curve and performs an inspection to identify temporal bins showing an excess of signal compared to the background. These temporal seeds will then be used to perform the imaging and mosaic analyses. 
The identification of time seeds first needs an estimation of the background. The background is estimated over the time window $[t_0 - 50 s, t_0+150s]$ with the removal of the on-source window, where $t_0$ is the external trigger time. By default the on-source window covers the range $[t_0 - 5 s, t_0+20s]$, but it can be adjusted by the user if there is evidence of signal outside that temporal range. This may be the case for long GRBs, where the brightest pulse can occur after the $t_0$. The user can inspect the raw count rate light curves binned with $[32, 64, 128, 256, 512, 1024, 2048, 4096, 8192]$ ms time bins. The search for temporal seeds is done using the same binning. For each choice of binning, the background temporal evolution is fit with a third-order polynomial and the best fit obtained with \texttt{emcee} adopting a Poissonian likelihood. The SNR of each time bin $j$ is calculated as:
\begin{equation}
    \operatorname{SNR}_j=\frac{C_j^{s u b}}{\sqrt{B_j+C_j}},
\end{equation}
where $C_j^{s u b}$ is the background-subtracted count rate, $C_j$ the bin count rate, and $B_j$ the background level coming from the previous fit. For visual inspection, the pipeline produces background-subtracted light curves where the time bins with SNR $>$ 3.5 are highlighted. The search for time seeds is restricted to the on-source time window. An example of the results of a time seeds search is given in \cref{fig:lc}, along with a plot showing the evolution in time of the spacecraft coordinates to identify slew periods.

If \textit{Swift} is in a stationary pointing mode, we perform a parallel analysis using the temporal seeds produced by \texttt{NITRATES}. The search of time seeds in \texttt{NITRATES} is explained in Sec. 7.3 of \cite{delaunay_nitrates}. The \texttt{NITRATES} time bins are ranked by SNR and the imaging search is performed on the top ten seeds. \texttt{NITRATES} does not produce time seeds if the on-source window overlaps with a period during which \textit{Swift} is slewing. In that case, only \texttt{BAT-GLIMPSE} temporal seeds are produced. 

The choice of the kind of analysis performed on each temporal bin depends on its duration. If the duration of the time bin is $<0.2$ s the imaging search is done, independently of whether the time bin overlaps with a slew phase. This approach is valid also during slew, since during a timescale $<0.2$ s the position of any transient in the BAT FOV would move by an apparent distance which is smaller than the width of the BAT point spread function \citep{slew_survey_copete}, equal to 22.5 arcmin FWHM\footnote{\url{https://heasarc.gsfc.nasa.gov/docs/software/lheasoft/help/batcelldetect.html}}. If instead the bin duration is $>0.2$ s and the bin overlaps with a slew phase, a mosaic is performed. We inspect the attitude file to determine when the spacecraft is slewing. We check if either the RA or Dec that the  spacecraft boresight is pointing towards changes by more than 1 arcminute between two subsequent times sampled in the attitude file. In the case of a slew, the bin is split into 0.2 s sub-bins and a sky image is produced for each of them. We then mosaic these images together over the relevant time period using the mosaic technique, as described in Sec. \ref{mosaic}.

If no source is found by imaging or mosaic, we carry out a refined search of seeds. For each temporal seed found before, defined by a duration $\Delta t$ and central time $t_c$, the refined search maximizes an objective function over a sample of time bins of duration $\hat{\Delta} t$ and central time $\hat{t}_c$, where $\hat{\Delta} t$ is sampled from a uniform distribution $\mathcal{U}[1/4 \Delta t, 4 \Delta t]$ and $\hat{t}_c$ is sampled from a uniform distribution $\mathcal{U}[t_c -\Delta t, t_c + \Delta t]$. The objective function assigns a score that is proportional to the background subtracted count rate.

\subsection{Imaging search}

The imaging analysis is performed using the \texttt{BatAnalysis} package \citep{2025ApJ...988...23P}, which provides a python interface with the HEASoft tools \texttt{batfftimage} and \texttt{batcelldetect}. For each time interval, a Detector Plane Image (DPI) is constructed in the 15–350 keV band. The DPI is then converted into a sky image, making use of the attitude file. Along with a sky image, a background standard deviation image and an SNR image are created. During the image creation, \texttt{batfftimage} uses the detection-optimized aperture map, with the option  \texttt{aperture="CALDB:DETECTION"}. The sky image has a minimum partial coding\footnote{The partial coding is defined as the fraction of the detector plane area illuminated by the source. It is maximum at the center of the BAT FOV (boresight) and declines towards the borders.} threshold of 0.01. A source is claimed as a confident detection if its position does not match any known cataloged source and if both SNR and centroid SNR are $>$ 5. The centroid SNR is the SNR in the center of the pixel that contains the position of the source. If multiple time seeds give a detection, \texttt{BAT-GLIMPSE} reports the position for the time seed that maximizes the imaging SNR.

\subsection{Mosaic search}
\label{mosaic}
The mosaic analysis combines sky images from multiple short time bins. We divide the entire temporal window into consecutive 0.2 s bins and use \texttt{BatAnalysis} to create sky images for each time bin.
The result is an object that contains a list of sky images, partial-coding images, and background standard-deviation images. Finally, the mosaiced image is created using multiple CPUs in parallel 
and converted into an healpix map with nside = 1024. A partial-coding threshold of 0.01 is adopted. When searching for new sources, we determine a source to be significant if SNR $>$ 5.5 and we classify it as a new $\gamma-$ray transient if the separation from the closest cataloged source is larger than one PSF full-width at half maximum.

\subsection{Real-time operations of \texttt{BAT-GLIMPSE}}

\begin{figure}
    \centering
    \includegraphics[width=1\linewidth]{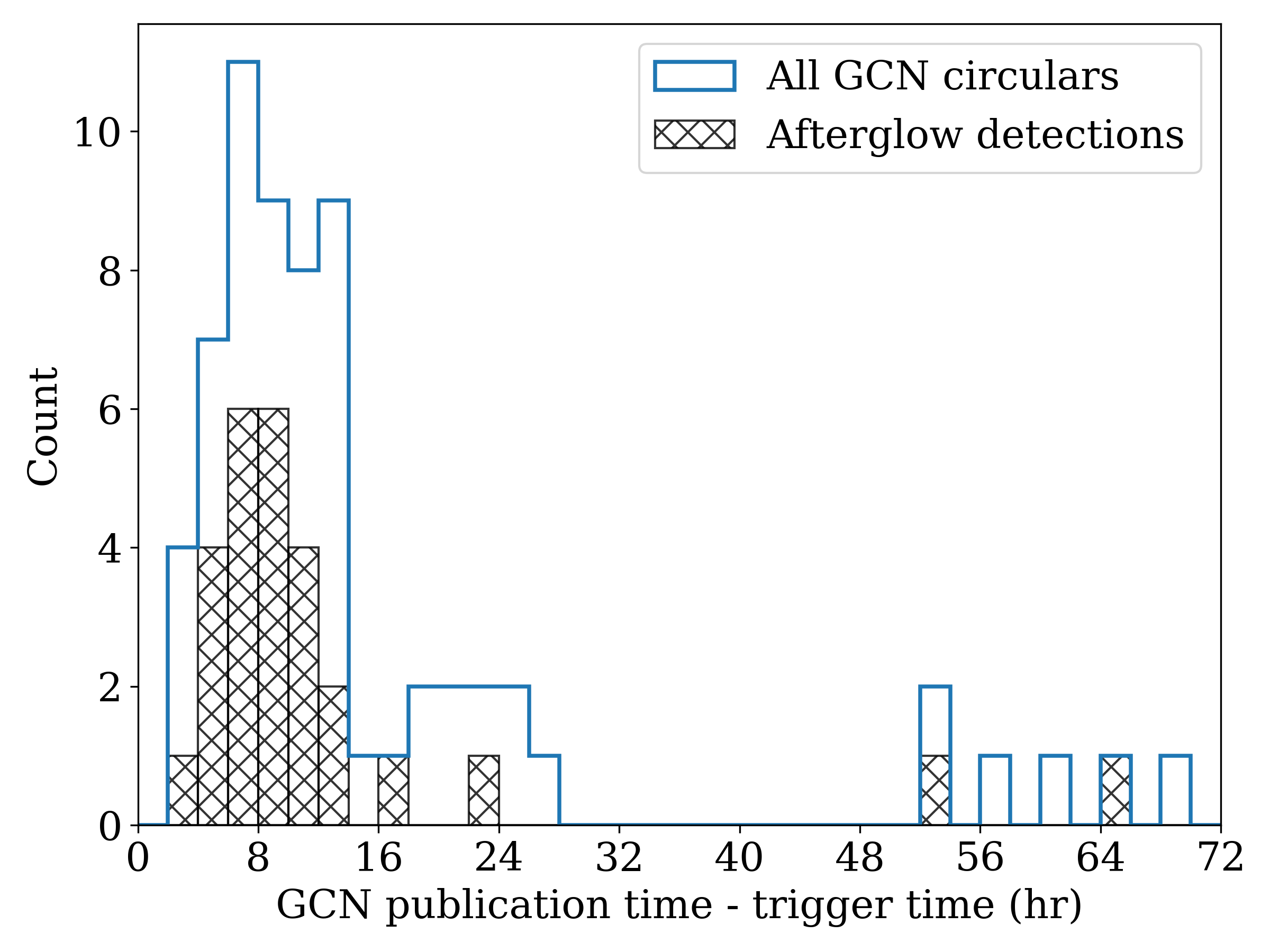}
    \caption{Histogram of the delay between the GUANO trigger time and the publication time of the GCN circulars containing GUANO arcminute localization. The hatched histogram is the sub-sample of GRBs whose afterglow was detected thanks to the GUANO position. Cases where the delay is larger than 72 hr are not included in the plot.}
    \label{fig:delay}
\end{figure}
\begin{figure*}
    \centering
    \includegraphics[width=1\linewidth]{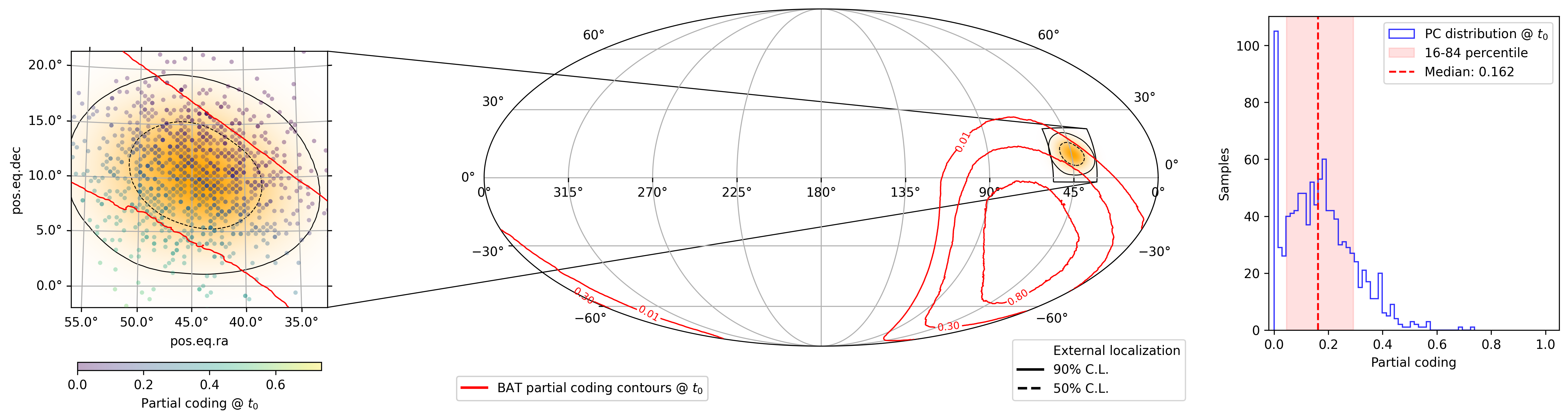}
    \caption{Example of diagnostic plots produced during the pre-run phase of \texttt{BAT-GLIMPSE}, in the case of \textit{Swift} in stationary pointing. The central panel shows the sky localization of the astronomical transient that triggered GUANO. The solid and dashed black lines indicate the 90$\%$ and $50\%$ credible levels. The red solid contours identify the 1$\%$, 30$\%$ and 80$\%$ partial coding fraction of BAT computed at the trigger time. The left panel contains an inset centered at the peak of the localization of the external trigger. The points are extracted from the sky localization of the transient. The right panel shows the partial coding probability distribution at $t_0$. These plots are for GRB 251013D \citep{2025GCN.42244....1D}.}
    \label{fig:imaging_map_1}
    \includegraphics[width=0.96\linewidth]{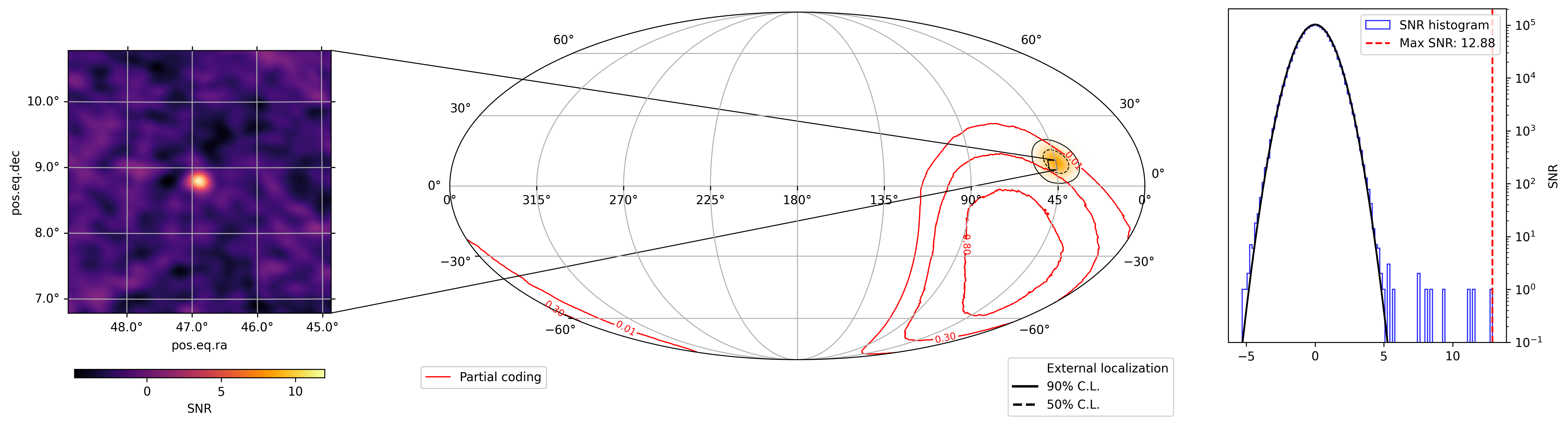}
    \caption{Example of summary plots produced after \texttt{BAT-GLIMPSE} detects and localizes the source, in the case of \textit{Swift} in stationary pointing. The central panel is the same of \cref{fig:imaging_map_1}. The left panel is a heat map of the SNR, while the right panel is an SNR histogram of the same sky image, along with a comparison with a Gaussian distribution centered at zero and standard deviation $\sigma =1$. The plots are for GRB 251013D \citep{2025GCN.42244....1D}.}
    \label{fig:imaging_map}
\end{figure*}
\begin{figure*}[]
    \centering
    \includegraphics[width=1\linewidth]{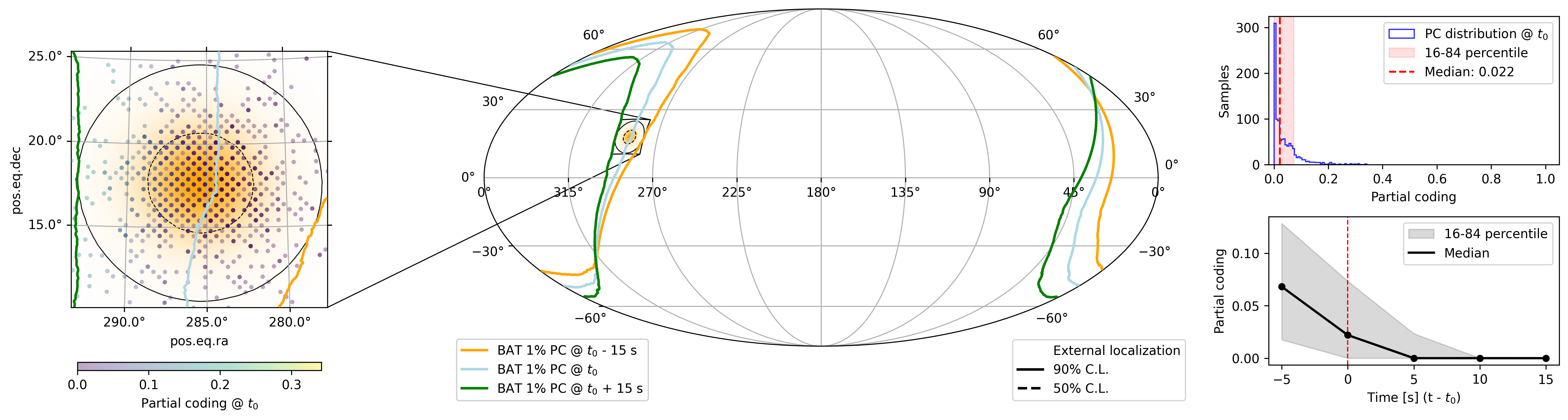}
    \caption{Example of diagnostic plots produced before the \texttt{BAT-GLIMPSE} run, in the case of \textit{Swift} slewing during at the external trigger time. The central panel shows the sky localization of the astronomical transient that triggered GUANO. The solid and dashed black lines indicate the 90$\%$ and $50\%$ credible levels. The green, light blue and orange solid contours identify the 1$\%$ partial coding fraction of BAT from $t_0$ - 15 s up to $t_0$ + 15 s, where $t_0$ is the external trigger time. The left panel contains an inset centered at the peak of the localization of the external trigger. The points are extracted from the sky localization of the transient. The right panels show the partial coding probability distribution at $t_0$ and its evolution in time (see text for more details). The plots are for GRB 220511A \citep{2022GCN.32023....1D}.}
    \label{fig:mosaic_1}    \includegraphics[width=1\linewidth]{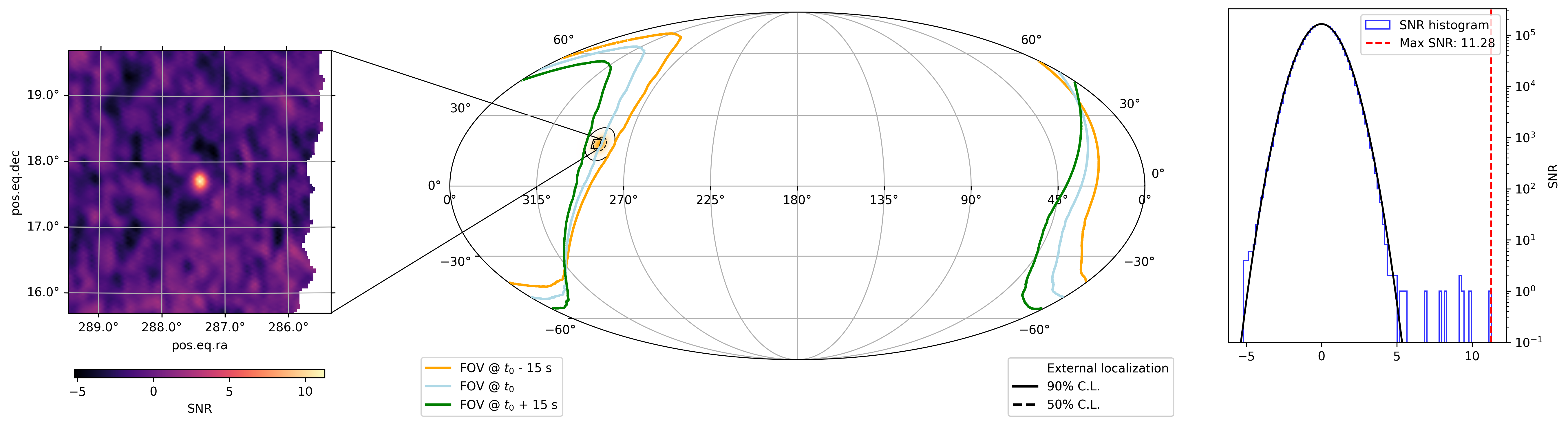}

    \caption{Example of summary plots produced after \texttt{BAT-GLIMPSE} detects and localizes the source, as \cref{fig:imaging_map} but in the case of a trigger time overlapping with a slew phase. The plots are for GRB 220511A \citep{2022GCN.32023....1D}.}
    \label{fig:mosaic_2}
\end{figure*}

\begin{figure}
    \centering
    \includegraphics[width=1\linewidth]{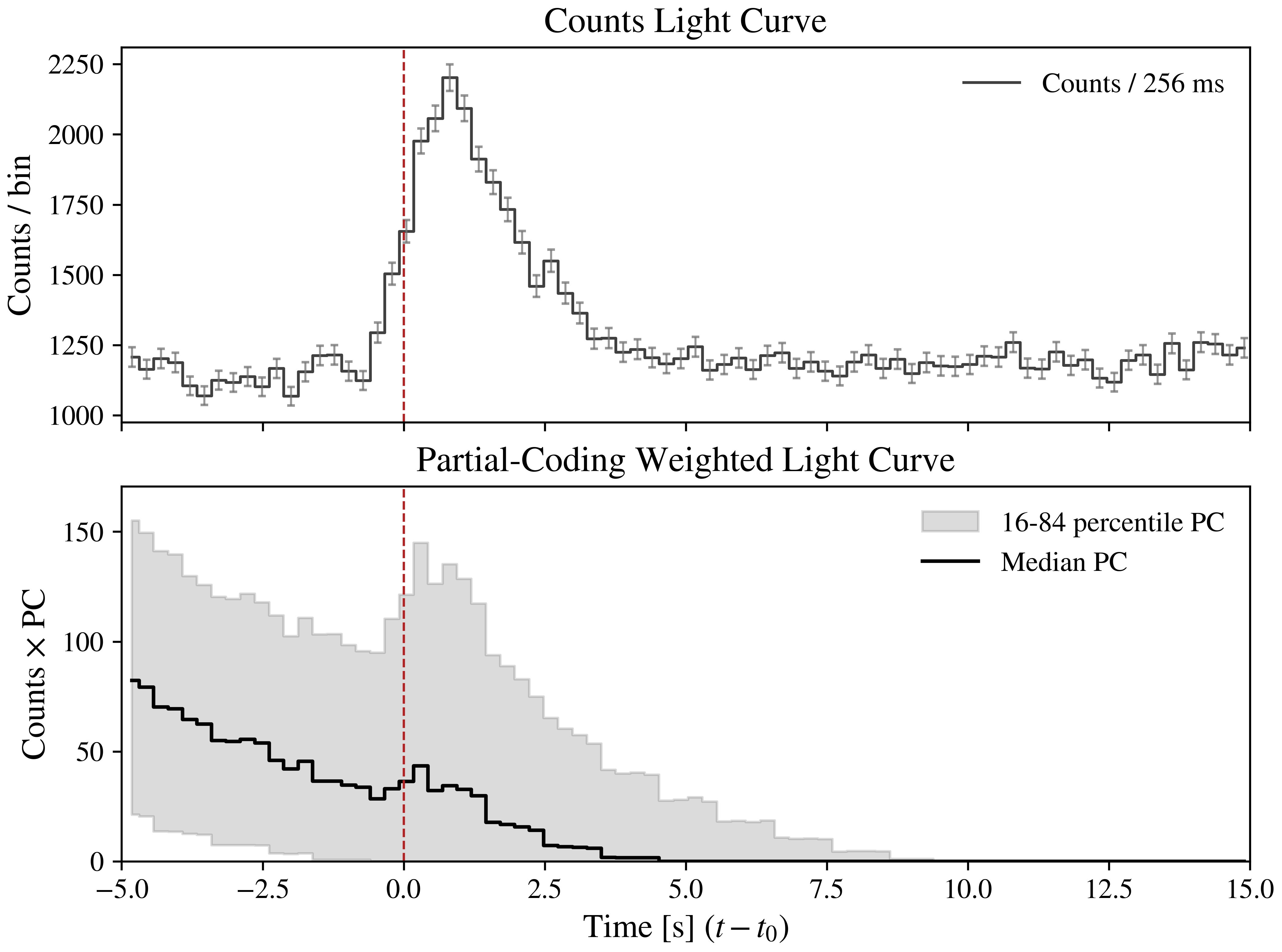}
    \caption{Top panel: raw 256 ms binned light curve of GRB 220511A (GCN 32013, \citealt{2022GCN.32023....1D}). Bottom panel: partial-coding weighted count light curve, which takes into account the apparent motion of the source across the BAT FOV. The distribution of partial coding obtained in the right panel of \cref{fig:mosaic_1} is used to produce the median (solid line) and 1$\sigma$ uncertainty band.}
    \label{fig:counts_pc}
\end{figure}

\texttt{BAT-GLIMPSE} became operational in real time at the beginning of 2025, working in synergy with \texttt{NITRATES}. The period of operation also covers the majority of the third part of the fourth observing run of the LIGO-Virgo-KAGRA (LVK) collaboration, improving the capability of \textit{Swift}-BAT to monitor, detect and localize the $\gamma-$ray counterpart of gravitational wave sources. As soon as GUANO data become available, \texttt{NITRATES} proceeds with data cleaning and generation of time seeds. Then \texttt{BAT-GLIMPSE} starts the analysis, first probing the time seeds provided by \texttt{NITRATES}, if available, and later performing a deeper search with larger number of trials. Currently \texttt{NITRATES} requires $\sim 30-45$ minutes to complete the analysis. \texttt{BAT-GLIMPSE} needs $\mathcal{O}$(1 minute) to run the imaging search and $\mathcal{O}$(10 minutes) to run the mosaic search, using multiple CPUs in parallel. As soon as \texttt{BAT-GLIMPSE} finds a potential candidate localization, the summary results are shared with the team of burst advocates in charge of the \textit{Swift}-BAT/GUANO targeted search.
At the time of writing, the \textit{Swift}-BAT/GUANO team has released seven circulars containing a GRB arcmin localization (2 short GRBs and 5 long GRBs) obtained with \texttt{BAT-GLIMPSE} \citep{2025GCN.39979....1R,2025GCN.41110....1R,2025GCN.41185....1R,2025GCN.42245....1D,2025GCN.42263....1R,2025GCN.43114....1R,2026GCN.43377....1R}. In all cases the \texttt{BAT-GLIMPSE} position was confirmed through cross-validation with \texttt{NITRATES} results. In two cases the detection of the afterglow was further confirmed by evidence of a fading source by \textit{Swift-}XRT \citep{2025GCN.43155....1O, 2026GCN.43426....1S}. 

\section{GRB sample selection} \label{grb_sample}

The availability on the ground of BAT time-tagged event data is made possible by the GUANO infrastructure, an automated pipeline that requests the downlink in response to external triggers of interest for the astronomical community. Inspecting the GCN archive \citep{GCN}, we searched for all the circulars published until March 2026, containing the detection of a GRB using GUANO, reporting a localization precision of $\mathcal{O}$(arcmin). These circulars fall into three categories: 1) a GCN reporting a detection using either imaging or mosaic techniques, 2) a GCN reporting a detection using the \texttt{NITRATES} analysis, 3) a GCN reporting a detection using the \texttt{NITRATES} analysis, and a comparison with the result obtained with imaging.

There is a total of 214 GUANO circulars reporting a GRB detection, of which 68 reporting a GRB with arcmin localization. All these sources are classified as GRBs, except for the SGR flare SGRJ1555-5402 (GCN 35022). In 16 cases ($\sim18\%$), the $\gamma-$ray source is detected during a spacecraft slew maneuver. For all the selected triggers, we download the BAT data using the GUANO API and we run the \texttt{BAT-GLIMPSE} analysis. In the case of GCN 31572, the event file is corrupted and the analysis cannot be completed. In the case of GCN 32506, the GUANO API does not give any data available. Therefore a complete \texttt{BAT-GLIMPSE} analysis was performed on 66 triggers.

Out of all the GRBs localized using GUANO data, the detection of an afterglow was confirmed for 26 of them. In \cref{tab:gua--} we provide the list of triggers along with relative GCN ID reporting the afterglow detection. Most of the time, the detection of the afterglow is provided by the evidence of fading of the X-ray light curve, as monitored by \textit{Swift-}XRT. In fewer cases, the afterglow confirmation comes also from detection in optical and radio bands. Furthermore, for an additional three cases, even if no afterglow was detected, the GUANO position was consistent with the error box provided by InterPlanetary Network (IPN) triangulation \citep{2013ApJS..207...38P,2020ApJ...895...40G,2022ApJS..259...34S,2025ApJS..278...60X}. Therefore the validity of the GUANO arcminute position has been confirmed independently for a total of 29 GRBs. 
\cref{fig:delay} gives an overview of the delay between the GRB trigger time and the publication time of the GCN circular, highlighting the sub-set of cases where the afterglow was detected. The majority of GUANO circulars that prompted the afterglow detection have been posted within $\sim$ 16 hr from the GRB trigger.

\section{Results} \label{results}

The results of \texttt{BAT-GLIMPSE} for all the GRBs collected in the GCN archive are reported in \cref{tab:gua--}. The table indicates the GCN identifier, the BAT sky coordinates reported in the GCN, the trigger time of the instrument that activated GUANO, the partial coding at the GRB position and trigger time, the $\sqrt{TS}$ given by \texttt{NITRATES} results when available, the SNR reported in the GCN (if only \texttt{NITRATES} results appear in the GCN, typically no corresponding imaging/mosaic SNR is reported) and the SNR found with \texttt{BAT-GLIMPSE} (absent if no source is identified). Moreover, the table indicates if the mosaic technique is used by \texttt{BAT-GLIMPSE} to detect the source and if the burst trigger time overlaps with any period during which the spacecraft is slewing. Finally the last column reports the angular separation between the position found by \texttt{BAT-GLIMPSE} and the one reported in the GCN.

\subsection{Diagnostic plots}

By default, the \texttt{BAT-GLIMPSE} pipeline generates a suite of diagnostic plots that aid in the interpretation of the results and in implementing user-defined adjustments aimed at improving search performance.

If the external trigger occurs while \textit{Swift} is in stationary pointing mode, a plot like the one of \cref{fig:imaging_map_1} is produced before the imaging analysis is performed. The plots reported in the figure are relative to GRB 251013D (GCN 42244,  \citealt{2025GCN.42244....1D}). The central panel shows a sky map with the localization of the external trigger, reporting 50$\%$ and $90\%$ credible levels, along with the contour levels of the BAT partial coding image, in order to identify the FOV. Using the external trigger sky localization map, we extract random positions following the corresponding probability distribution. The extracted positions are shown in the inset on the left of \cref{fig:imaging_map_1}, where each point has a color coded according to the value of the partial coding. On the right panel of \cref{fig:imaging_map_1} we show the posterior distribution of the partial coding, with a vertical red bar indicating the 1$\sigma$ equivalent uncertainty. This last plot helps the user in predicting the chance that the source is inside the BAT FOV and is detectable through imaging, before the actual analysis is performed.

If \texttt{BAT-GLIMPSE} reports a significant detection of a source, a plot like \cref{fig:imaging_map} is produced. The central panel is the same of \cref{fig:imaging_map_1}. The left panel shows a color map of the SNR image projected onto a healpix map with nside = 512. The plot is centered at the position found by \texttt{BAT-GLIMPSE}. The right panel reports the SNR histogram along with a black solid curve indicating the expected Gaussian distribution in the null hypothesis of the absence of a real signal. If multiple time seeds have a corresponding detection using imaging, both the sky image map and the SNR histogram plot are produced selecting the time window that maximizes the SNR in the \texttt{BAT-GLIMPSE} search.

If the GRB trigger time overlaps with a period of slew, a plot like the one of \cref{fig:mosaic_1} is produced before the beginning of the analysis. The figure shows results relative to GRB 220511A (GCN 32013, \citealt{2022GCN.32023....1D}). Similarly to \cref{fig:imaging_map_1}, the central panel shows a sky map with the localization of the external trigger, along with the contour at 1$\%$ partial coding fraction at $t_0$ - 15 s, $t_0$, and $t_0$ + 15 s. The panel on the left contains an inset centered at the peak of the probability map of the external trigger, along with a collection of positions sampled by the sky probability distribution, colored by the value of partial coding. On the right, the upper panel is analogous to the one of \cref{fig:imaging_map_1}, computed at $t_0$, while the lower panel is the time evolution of the partial coding averaged over the samples extracted from the external trigger localization map. At each time, from $t_0-5$ s up to $t_0$+15 s, we compute the distribution of partial coding and report in the plot the median with a solid line and the 1$\sigma$ equivalent uncertainty with a gray band. In the specific case of \cref{fig:mosaic_1}, the panel shows that at $t_0$ the GRB is likely at a partial coding fraction $<10\%$ and already at $t_0$+5 s the source is likely outside the coded mask FOV. This product can help the user in making an ad-hoc choice of the time window for the mosaic, avoiding to include those times where the source is outside the BAT FOV. To help even further in this task, \texttt{BAT-GLIMPSE} also includes the plot of the count rate light curve (by default binned at 256 ms) re-weighted by the partial coding, as shown in \cref{fig:counts_pc}. The solid black line is the median value of the re-weighted counts, while the gray band is the 1$\sigma$ equivalent uncertainty. The optimal temporal window for the mosaic results from a trade-off between maximizing the source count rate and maximizing the partial coding fraction, as both factors contribute to the detection significance. In case \texttt{BAT-GLIMPSE} does not detect the source autonomously, the user can make use of this diagnostic plot to chose an ad-hoc temporal window.

\begin{figure}
    \centering
    \includegraphics[width=0.95\linewidth]{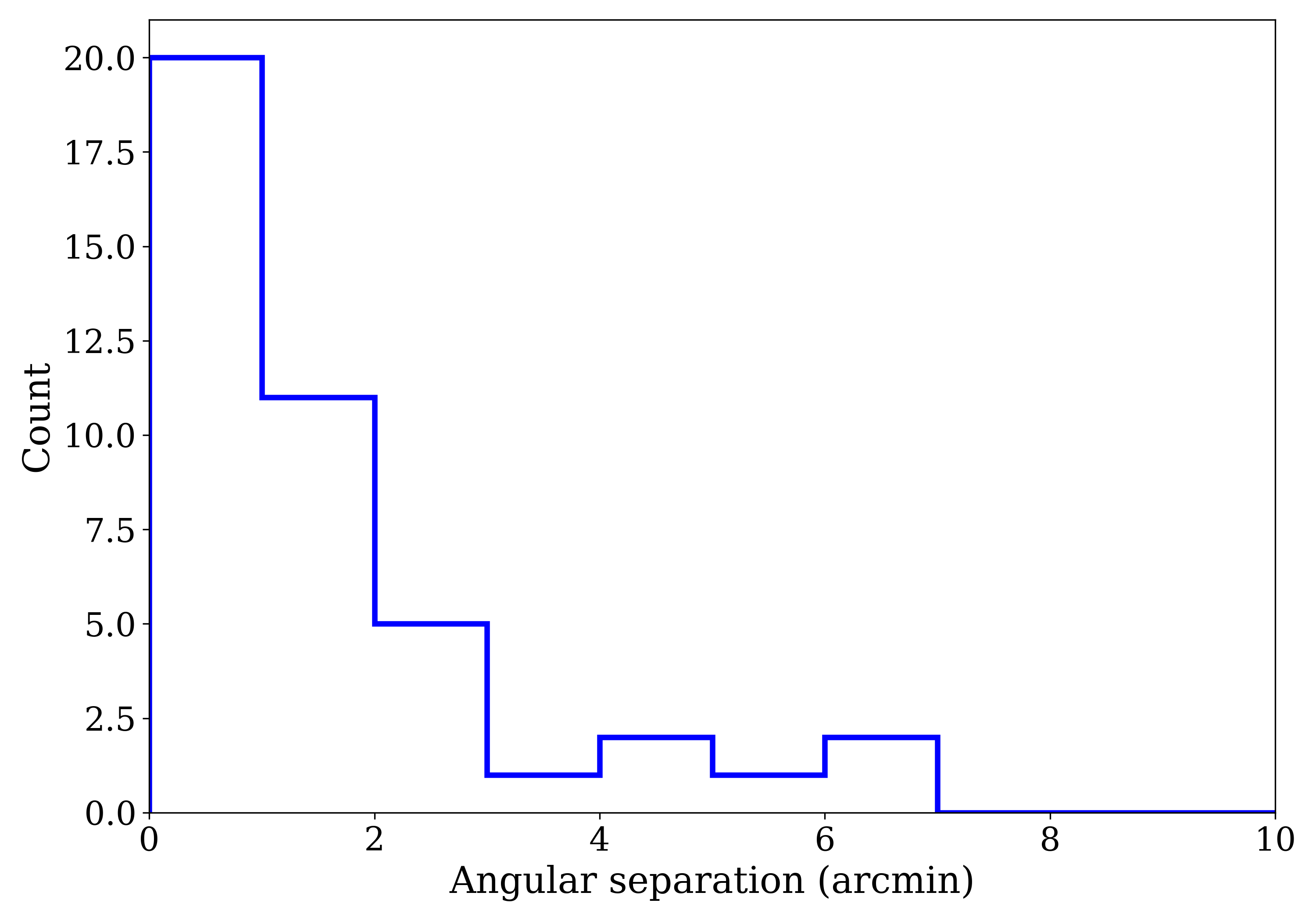}
    \caption{Distribution of the angular separation between the GRB position found by \texttt{BAT-GLIMPSE} and the one reported in the GUANO GCNs.}
    \label{fig:snr_hist}
\end{figure}

\begin{figure}
    \centering
    \includegraphics[width=1\linewidth]{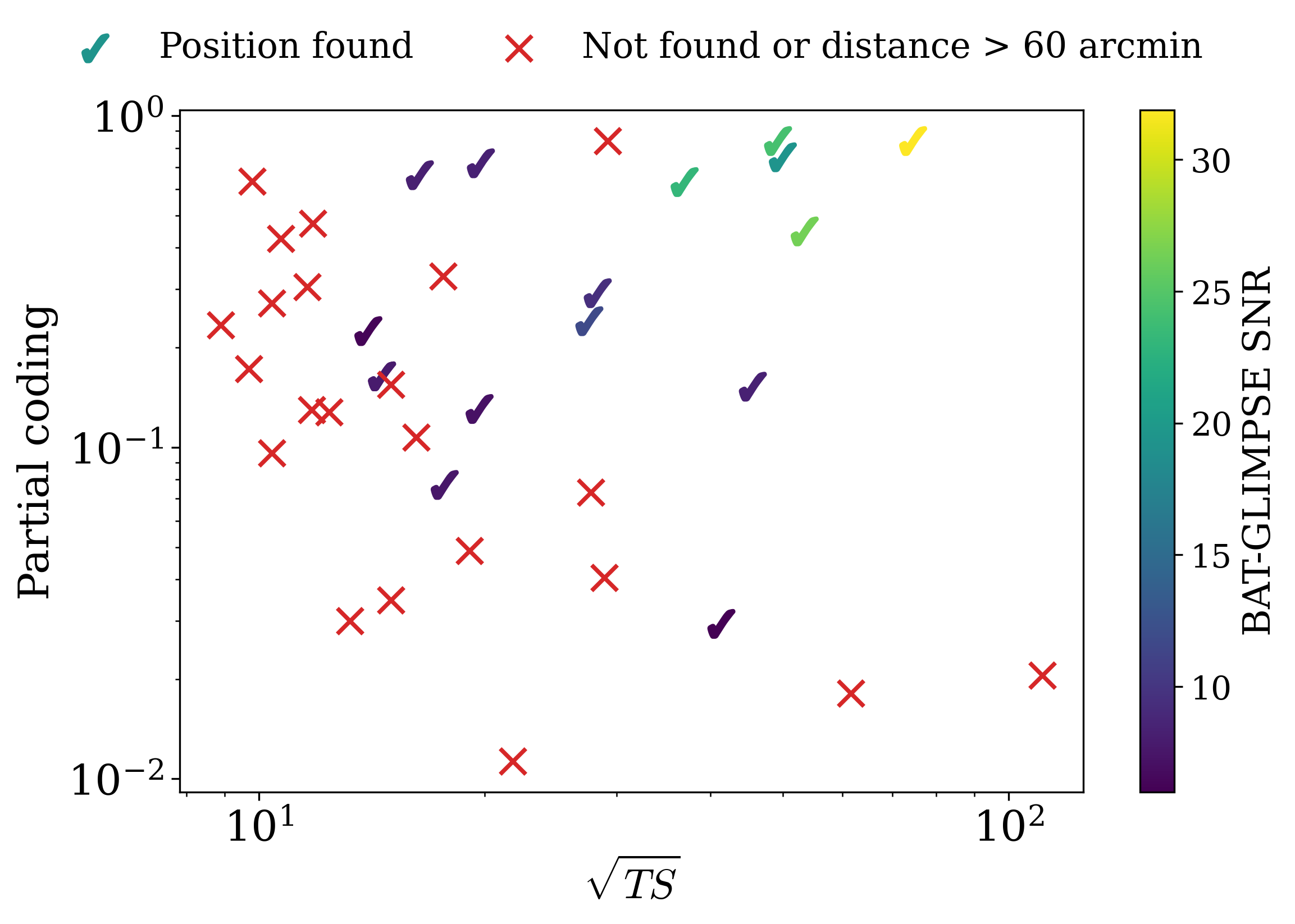}
    \caption{Distribution in the partial coding - $\sqrt{TS}$ plane of all the GRBs localized by \texttt{NITRATES} with arcmin precision. The points indicated with a cross are GRBs not detected by \texttt{BAT-GLIMPSE}, while the $\checkmark$ symbol indicates GRBs detected both by \texttt{NITRATES} and \texttt{BAT-GLIMPSE}. The color bar indicates the SNR measured by \texttt{BAT-GLIMPSE}.}
    \label{fig:ts_vs_pc}
\end{figure}

\subsection{Validation of \texttt{BAT-GLIMPSE} and comparison with \texttt{NITRATES}}

Out of the 66 GRBs analyzed, for 43 of them \texttt{BAT-GLIMPSE} detects the source and the position is consistent with the one reported in the GCN. Whenever \texttt{BAT-GLIMPSE} detects the source, the angular separation is $\lesssim 5$ arcmin. The distribution of angular separation is shown in \cref{fig:snr_hist}. We exclude from the plot GRB 240317B (GCN 35948, \citealt{2024GCN.35948....1D}), since in a later circular (GCN 35965, \citealt{2024GCN.35965....1D}) the authors report that \textit{Swift} was in safe point mode and had an unstable pointing, implying a likely underestimation of the position uncertainty. 

In \cref{tab:gua--}, an angular separation $>60$ arcmin indicates that the source found by \texttt{BAT-GLIMPSE} is likely an artifact (noise fluctuation). This happens only when the imaging search is used and the source is found with an SNR $<$ 7. In the range $6<\text{SNR}<7$ \texttt{BAT-GLIMPSE} recovers the correct position of the source only in 2/5 cases. We therefore recommend treating as highly reliable only the positions found with an imaging SNR $>$ 7. In the collection of the GCN, the position is reported mentioning \texttt{NITRATES} for 31 GRBs, and for 15 of these \texttt{BAT-GLIMPSE} recovers the same position using imaging search. The \texttt{NITRATES} pipeline allows to perform a search of signals more sensitive than the imaging search. Within \texttt{NITRATES}, the statistical significance of a candidate detection is quantified using the likelihood-ratio Test Statistic (TS). The definition of TS can be found in \citealt{delaunay_nitrates}. A value of $\sqrt{TS}\sim 8$ corresponds to a false alarm rate comparable to the GRB rate and hence is used to claim a confident detection. Given the higher sensitivity of \texttt{NITRATES} compared to \texttt{BAT-GLIMPSE}, we expect that for low values of $\sqrt{TS}$ the \texttt{BAT-GLIMPSE} is not able to recover the GRB position. This is illustrated in \cref{fig:ts_vs_pc}, which shows the distribution of \texttt{NITRATES} and \texttt{BAT-GLIMPSE} detections in the partial coding - $\sqrt{TS}$ plane. The color bar indicates the \texttt{BAT-GLIMPSE} SNR. All the sources reported with a $\checkmark$ are detected both by \texttt{NITRATES} and \texttt{BAT-GLIMPSE}.

Out of the 16 GRBs occurring during a slew phase, \texttt{BAT-GLIMPSE} recovers the arcmin position for 8 of them using mosaic. For 6 cases \texttt{BAT-GLIMPSE} recovers the position using only the imaging search on a temporal window shorter than 0.2 s. This happens when the source is bright enough that a single 0.2 s image contains enough counts to claim a detection. For only 2 cases the source is not detected with neither mosaic or imaging and in both cases the GRB is located at a partial coding $<10\%$. 

\subsection{Upper limits from non-detection}

\begin{figure*}[]
    \centering
\includegraphics[width=1\linewidth]{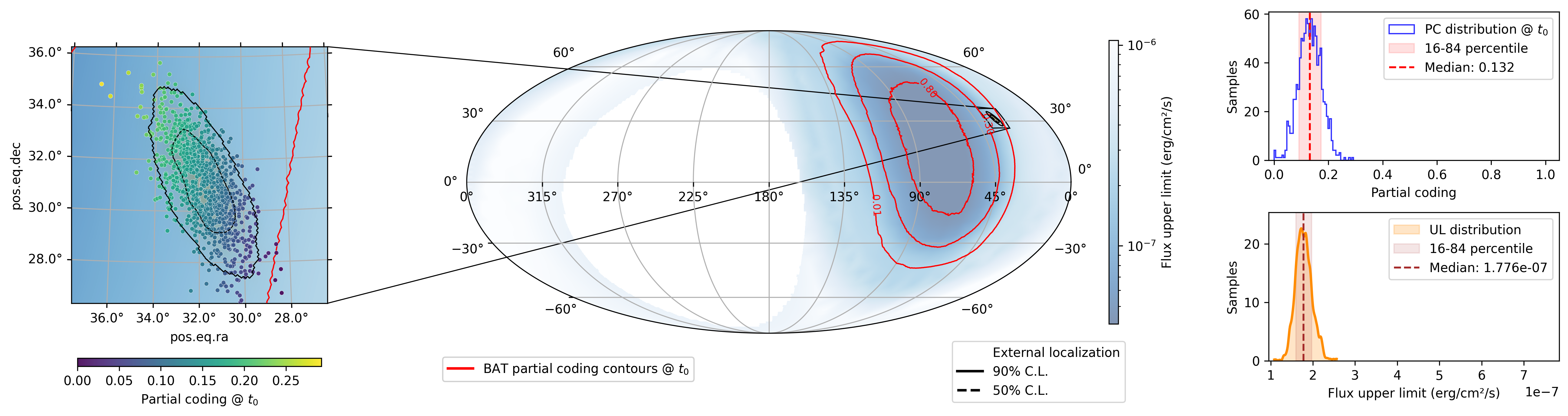}
    \caption{Diagnostic plot for the GW candidate S250331o \citep{2025GCN.39953....1L}. \texttt{BAT-GLIMPSE} can be used to derive the distribution of partial coding to determine the probability that the GW candidate is inside the BAT FOV at the trigger time. The blue color map indicates the flux upper limits derived by \texttt{NITRATES}. On the right the two panels show the distribution of partial coding and flux upper limits averaged over the sky probability distribution of the GW candidate. The GW sky localization map is taken from \cite{2025GCN.39986....1L}.}
    \label{fig:gw}
\end{figure*}

\begin{figure*}[]
    \centering
\includegraphics[width=1\linewidth]{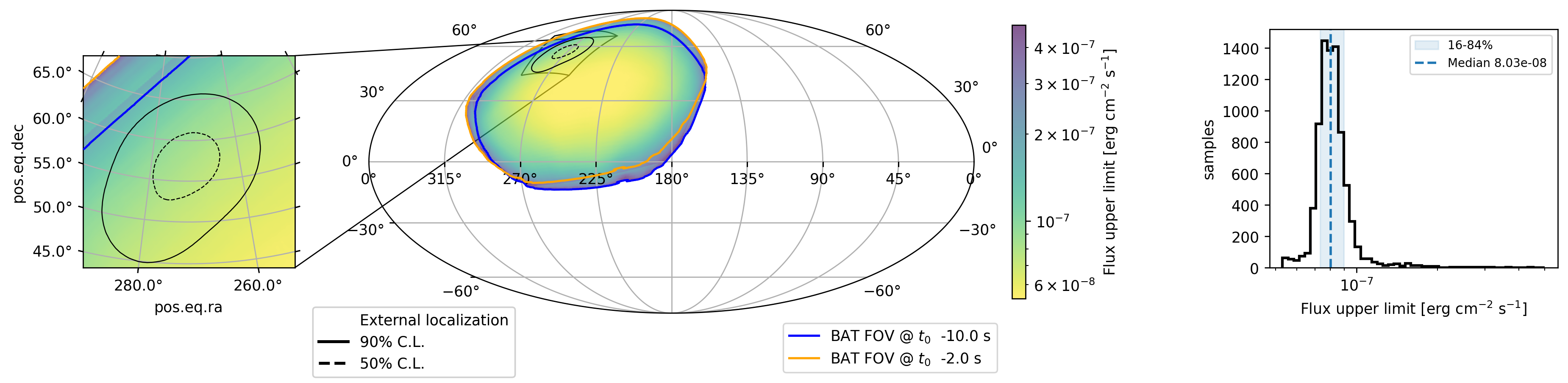}
    \caption{Flux upper limit map from a mosaiced image. The blue and orange solid lines indicate the BAT FOV at the beginning and end of the time window. The right panel shows the distribution of the upper limit obtained with positions injected according to the sky probability distribution of the external trigger. The plots are relative to GRB 210421B \citep{2021GCN.29877....1D}, in the time interval prior to the GRB trigger, $[t_0 -10 \, s, t_0 -2 \, s]$. }
    \label{fig:ul_slew}
\end{figure*}

In case of non-detection, \texttt{BAT-GLIMPSE} produces flux upper limits maps, in healpix fits format. If the trigger does not overlap with slewing, the upper limits are computed using the background estimation and sensitivity of
\texttt{NITRATES}, following the procedure used by \cite{2025ApJ...980..207R}. An example of flux upper limit map is shown in \cref{fig:gw} for the GW candidate S250331o \citep{2025GCN.39953....1L,2025GCN.39986....1L}, along with the upper limit distribution averaged over the injections that follow the sky probability map of the external trigger, in the right panel. The flux upper limits are obtained in the energy range $[15,350]$ keV and in a time window $[t_0 -20 \, s, t_0 + 20 \, s]$, adopting the "normal" spectrum defined in \cite{2025ApJ...980..207R} and for a temporal bin of 1024 ms.

In addition to the partial coding distribution, this plot determines the probability $ p_{\rm inFOV}$ that the external candidate is inside the BAT FOV at the trigger time, corresponding to the number of sky coordinate injections that are above a minimum partial coding $\rm pc_{min}$, namely
\begin{equation}
    p_{\rm inFOV} =\rm N_{inj}( PC>\rm pc_{min})/N_{inj},
\end{equation}
where $\rm pc_{min}$ can be set to 0.01.
This information is relevant to interpret how constraining non-detection BAT upper limits are
(\citealt{2024ApJ...970L..20R,2025ApJ...980..207R,2025GCN.40109....1W}).

If the trigger does overlap with slewing, and a mosaic approach is necessary, \texttt{BAT-GLIMPSE} produces a flux upper limit map as well, using a time window specified by the user. The background standard deviation map of the mosaiced image is used to obtain the upper limit map in units of count rate. 
Calling $C_{e_1-e_2}(p)$ the count rate of the  pixel $p$ in the $[e_1,e_2]$ energy range, the flux upper limit map at $n_{\sigma}$ confidence level (by default set to 5$\sigma$)  is:
\begin{equation}
    F_{\text{UL}}^{e_1-e_2} (p)= K (p) n_{\sigma}C^{e_1-e_2}(p),
\end{equation}
where $K(p)$ is the function that converts counts to flux for each pixel. The $K(p)$ function is defined as
\begin{equation}
    K(p)=\frac{\int E\phi(E)  d E}{ \int \phi(E) A_{eff}(E, p) d E},
\end{equation}
where $A_{eff}$ is the effective area and $\phi(E)$ the assumed photon spectrum. $A_{eff}$ is derived using an interpolation of the detector response matrix (DRM) on a grid of sky positions sampled inside the FOV.
As explained in sec. 5.7.7 of the BAT software guide\footnote{\url{https://swift.gsfc.nasa.gov/analysis/swiftbat.pdf}}, it is an acceptable approximation to use a single DRM for the whole slew period under the condition that the slew does not span more than 15 deg. Since the \textit{Swift} slew rate can be maximum $\sim$1 deg/s, it is legitimate\footnote{This approach is acceptable also because we are not performing a spectral fit to obtain physical spectral parameters, as done in section 4.4 of \cite{2025ApJ...988...23P}.} to use a single DRM for flux upper limits relative to a time window shorter than $\sim 15$ sec. An example of a [15-350] keV flux upper limit map during a slew is shown in \cref{fig:ul_slew}, relative to GRB 210421B \citep{2021GCN.29877....1D}. Even if the GRB was detected by \texttt{BAT-GLIMPSE} with mosaic, the time interval for the upper limit map is chosen intentionally before the GRB trigger time, to ensure absence of signal. The figure shows with a color map the value of the flux upper limit for each position of the sky spanned by the BAT FOV during the slew.  The right panel shows the distribution of the upper limit averaging over the sky probability distribution of the external trigger. The spectrum adopted here is the same one used for \cref{fig:gw}.

\section{GRB detection sensitivity of \texttt{BAT-GLIMPSE} + \texttt{NITRATES} } \label{sensitivity}

Considering only the GCN circulars published during the period of time where \texttt{NITRATES} started to run at full regime (i.e., with $100\%$ efficiency in processing all GUANO triggers\footnote{However, at the beginning GUANO was not yet fully processing triggers during slews}), we find an overall increase in the \textit{Swift-}BAT GRB detection rate of +70.3$\%$ with respect to the onboard trigger rate\footnote{The average in the period 2020-2025 corresponds to 63.3 GRBs / yr, as it can be checked from the \textit{Swift-}BAT GRB catalog: \url{https://swift.gsfc.nasa.gov/archive/grb_table/stats/}}. Regarding just the 
detection rate of GRBs localized at arcminute precision, we find an increase of $\sim$+17.4$\%$ with respect to the onboard trigger rate. We note that the subset of \texttt{NITRATES} detections not localized at the arcminute precision include GRBs occurring both inside and outside the FOV. 

\subsection{Gain in GRB detection rate by \texttt{NITRATES} alone}
\begin{figure}[t]
    \centering
    \includegraphics[width=1\linewidth]{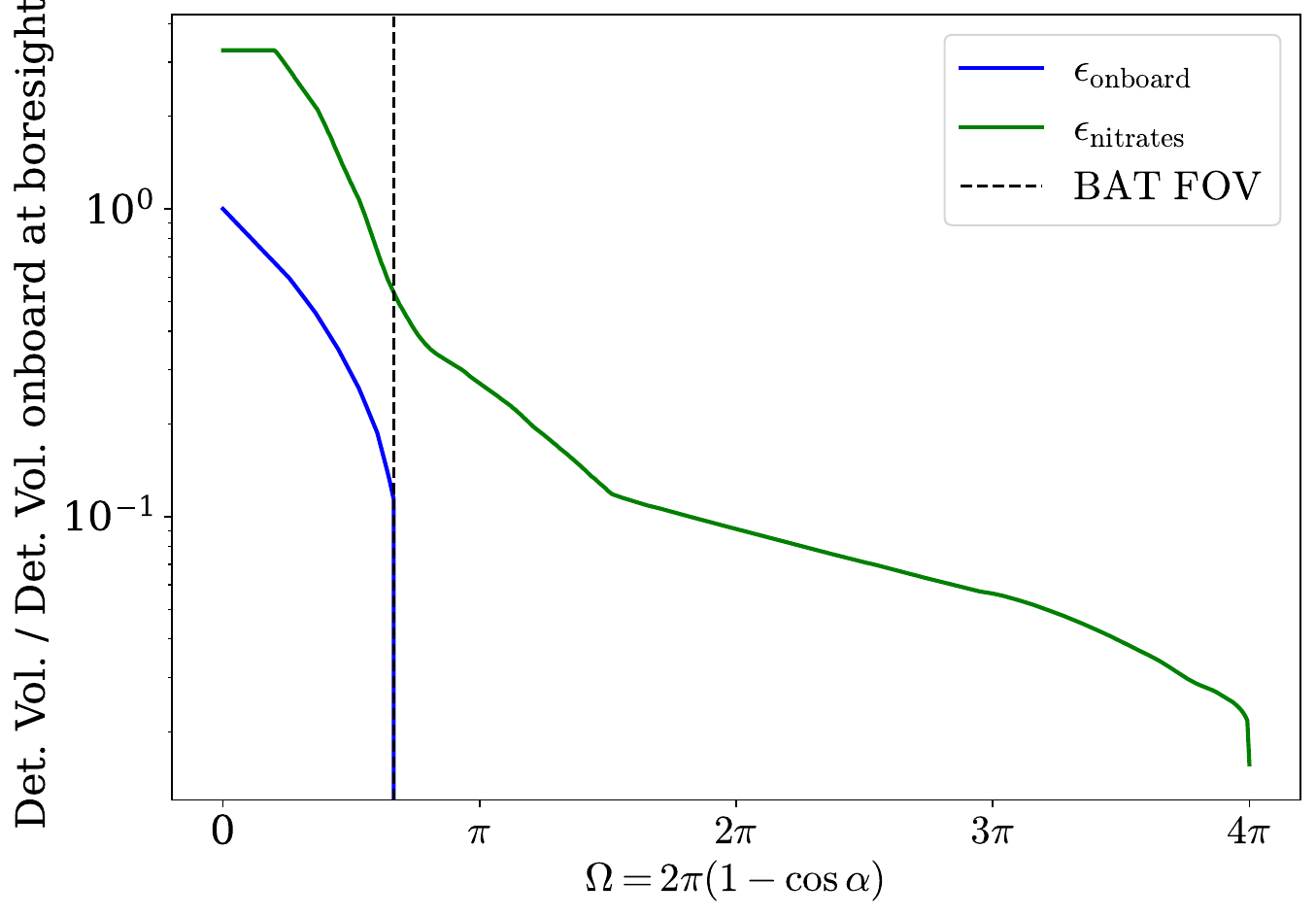}
    \caption{Detectable volume of \texttt{NITRATES} with respect to BAT onboard imaging as a function of the angular distance $\alpha$ between the source and the BAT boresight, averaged over the azimuthal angle. The solid angle enclosed between the source and the boresight is indicated with $\Omega$. Each curve is normalized by the detectable volume relative to a source locate the center of the FOV. A source with temporal and spectral properties like the ones of GRB 170817A is assumed \citep{2017ApJ...848L..14G}.}
    \label{fig:nit_sens}
\end{figure}

\cite{delaunay_nitrates} derive a factor $1.37$ for the ratio between rates SNR vs imaging SNR. This corresponds to a gain $\sim$ 1.6 in detectable volume under the assumption that it scales as the sensitivity to the power 3/2. The equality gain $=$ 1.6 holds in the case of a uniform-in-volume distribution of sources, an approximation that is valid for local sources, for instance if the external trigger is a GW from a neutron star merger detected within the current LVK horizon. The actual gain in sensitivity of \texttt{NITRATES}, however, is only in rough approximation equal to the ratio between rates SNR vs imaging SNR. \cref{fig:nit_sens} shows the gain in sensitivity of \texttt{NITRATES} obtained simulating a signal like GRB 170817A \citep{2017ApJ...848L..14G} distributed uniformly across the sky. The sensitivity gain is shown in terms of the ratio between the detectable volume of \texttt{NITRATES} and the detectable volume for the same source located at the BAT boresight, as a function of the angular distance $\alpha$ from the BAT boresight. The result is averaged over the azimuthal angle. A threshold of $\sqrt{TS}=7.5$ is used to define a \texttt{NITRATES} detection, corresponding to a false alarm rate of 1 / hr \citep{delaunay_nitrates}. Focusing only on the sources inside the BAT FOV, the average gain in detectable volume can be expressed as
\begin{equation}
\label{g_in}
    g_{\rm inFOV} = \dfrac{\int_{\rm inFOV}\epsilon_{\rm nitrates}(\Omega)d \Omega}{\int_{\rm inFOV}\epsilon_{\rm onboard}(\Omega)d \Omega}\simeq 4.3,
\end{equation}
where $\epsilon_{\rm nitrates}(\Omega)$ and $\epsilon_{\rm onboard}(\Omega)$ are the detectable volumes of \texttt{NITRATES} and BAT onboard trigger, respectively. The integral is restricted to the BAT FOV only. The extra gain in detection rate given by the sources outside the FOV detectable only by \texttt{NITRATES} is given by
\begin{equation}
\label{g_out}
\begin{gathered}
    g_{\rm outFOV} = \\
    \left(\frac{4\pi}{\Omega_{\rm FOV}}-1\right)\dfrac{\int_{\rm outFOV}\epsilon_{\rm nitrates}(\Omega)d \Omega}{\int_{\rm inFOV}\epsilon_{\rm onboard}(\Omega)d \Omega} \simeq 0.41,
\end{gathered}
\end{equation}
where $\Omega_{\rm FOV}$ is the solid angle area covered by the BAT FOV. The integral at the numerator is now restricted to the area outside the BAT FOV. The factor $\left(\frac{4\pi}{\Omega_{\rm FOV}}-1\right)$ is the ratio of GRBs occurring outside the BAT FOV over those inside.
Overall, \texttt{NITRATES} gives access to an extra amount of GRB detections equal to $(g_{\rm inFOV}+g_{\rm outFOV})R_{\rm onboard}\simeq 4.71\times R_{\rm onboard}$, with $R_{\rm onboard}$ the onboard BAT GRB average trigger rate. The reason why the actual measured increase of GRB detection rate given by \texttt{NITRATES} is smaller (+70.3$\%$ reported at the beginning of this section) is mostly due to the limited sensitivity of the external instruments that trigger GUANO and allow the \texttt{NITRATES} analysis. Secondary causes are related to the fact that the values predicted by \cref{g_in,g_out} are valid using GRB 170817A as benchmark, and also because the gain in detection volume is equal to the gain in detection rate only for a population of sources distributed uniformly in volume. 

\subsection{Gain in GRB detection rate by \texttt{BAT-GLIMPSE}+\texttt{NITRATES}}

Since \texttt{NITRATES} and \texttt{BAT-GLIMPSE} operate only if triggered by GUANO, namely if an external instrument detected a GRB, the effective recovery efficiency depends on the sensitivity of the external triggering instrument. The net gain introduced by \texttt{BAT-GLIMPSE} is only relative to the positions recovered during slew, since during pointing the sensitivity of \texttt{BAT-GLIMPSE} is more limited than the \texttt{NITRATES} one. Let us define $\epsilon_{\rm ext}$ the ratio between the volume detectable by the external instrument and the volume detectable with BAT onboard imaging. Correspondingly, we define $\epsilon_{\rm glimpse}$ as the ratio between the volume detectable by \texttt{BAT-GLIMPSE} and the volume detectable with BAT onboard imaging. Considering only sources inside the BAT FOV, we can write the overall fractional gain in detection rate obtained by the combination of \texttt{NITRATES} and \texttt{BAT-GLIMPSE} as:
\begin{equation}
\begin{gathered}
\rho_{\rm inFOV} =\frac{R_{\text {nitrates+glimpse }}}{R_{\mathrm{BAT}}}\Bigg|_{\rm inFOV}=\\
 =\min(g_{\rm inFOV},\epsilon_{\rm ext}) + f_s \min(\epsilon_{\rm glimpse},\epsilon_{\rm ext}) ,
\end{gathered}
\end{equation}
where $f_s$ is the fraction of time spent by \textit{Swift} slewing, and $R_{\rm BAT}$ is the GRB detection rate of BAT considering only the onboard triggering algorithm. If we instead consider also the GRBs outside the FOV found by \texttt{NITRATES}, the fractional gain in detection rate is 
\begin{equation}
\label{eq:rho_eq}
\begin{gathered}
\rho_{\rm inFOV+outFOV} = \\
=\frac{R_{\text {nitrates+glimpse }}}{R_{\mathrm{BAT}}}\Bigg|_{\rm inFOV+outFOV}=\\
=\rho_{\rm inFOV} + \min\left(g_{\rm outFOV}, \left(\frac{4\pi}{\Omega_{\rm FOV}}-1\right)\epsilon_{\rm ext}\right).
\end{gathered}
\end{equation}
Since external triggering instruments are not substantially less sensitive than BAT, we can conservatively assume $\left(\frac{4\pi}{\Omega_{\rm FOV}}-1\right)\epsilon_{\rm ext}>1$. Therefore, since $g_{\rm outFOV}<1$, we can write
\begin{equation}
\begin{gathered}
\rho_{\rm inFOV+outFOV} = \\
\min(g_{\rm inFOV},\epsilon_{\rm ext}) + f_s \min(\epsilon_{\rm glimpse},\epsilon_{\rm ext}) 
\\+ g_{\rm outFOV}.
\end{gathered}
\end{equation}
To be noticed that $R_{\rm BAT}$ is generally slightly smaller than the \texttt{BAT-GLIMPSE} recovery rate achievable offline, as demonstrated by all the additional GRBs recovered by \texttt{BAT-GLIMPSE} using imaging, not during slews (15 GRBs recovered in 4 yr corresponding to $\sim+6\%$ in addition to the average 63.3 onboard detections / yr). 

As mentioned before, \cref{eq:rho_eq} are valid if there is a linear scaling between the gain in detection rate and the gain in detectable volume, namely only for sources distributed uniformly in space. For a population with a density that increases (decreases) with distance, an increase of detectable volume $\Delta V$ has a correspondent increase in rate that scales faster (slower) than linearly with $\Delta V$. GRB population studies indicate densities that peak at redshifts $z\sim 1-2$ for the short-duration ones \citep{2016A&A...594A..84G} and $z\sim 3-4$ for the long-duration ones \citep{2022ApJ...932...10G}. If most of the sources detected by BAT are closer than the peak of the density distribution, \cref{eq:rho_eq} is a conservative estimate. However, the exact result is a non-trivial interplay between the specific evolution in distance of the density and the luminosity function of the source. Regardless, \cref{eq:rho_eq} remains a solid benchmark to assess the sensitivity gain of the two searches, for all the GRBs occurring both inside and the BAT FOV.

\begin{figure}[t!]
    \centering
    \includegraphics[width=\linewidth]{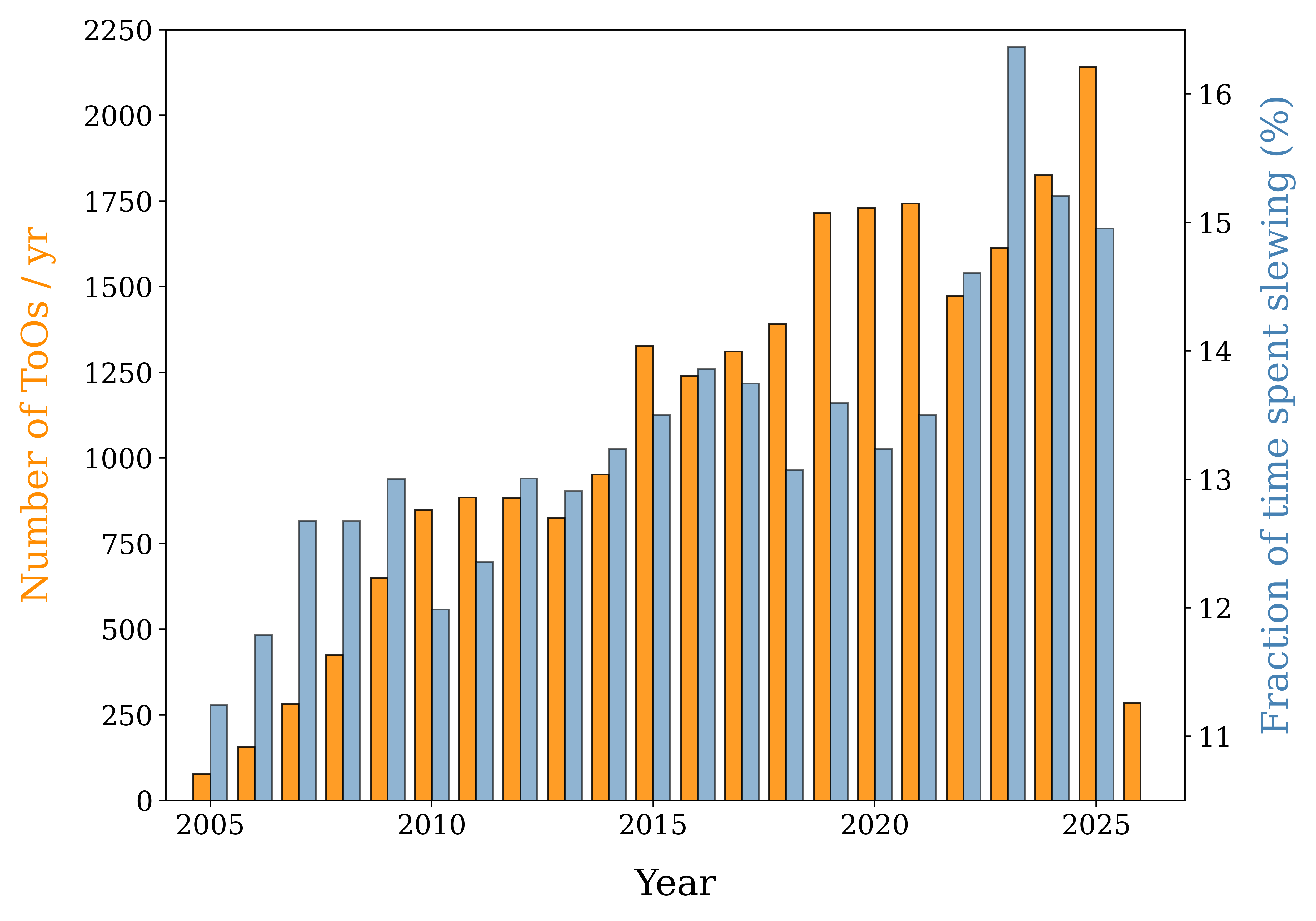}
    \caption{Time evolution of the number of ToO requests / yr (left y axis) and the fraction of time spent by \textit{Swift} slewing (right y axis), from the lauch time until end of 2025.}
    \label{fig:slew_frac}
\end{figure}

To compute the fraction $f_s$ in \cref{eq:rho_eq}, we used the \texttt{swifttools} package\footnote{\url{https://www.swift.psu.edu/too_api/}}, and we collected all the observation data from 2005 to 2025. The evolution of $f_s$ is shown in \cref{fig:slew_frac}, right y axis, along with the number of ToO requests submitted / yr, left y axis. It is apparent that, though there is a correlation between $f_s$ and the ToO submission rate, the proportionality constant is $\ll1$, since the scheduling of ToO observations is optimized to minimize the time spent slewing from one target to the next one. In the last years $f_s$ reached an average value $\sim15\%$. 

Evaluating the minimum and maximum values of \cref{eq:rho_eq}, we obtain a range of fractional gain 
\begin{equation}
\rho_{\rm min} < \rho_{\rm inFOV+ \rm outFOV}  < \rho_{\max}, 
\end{equation}
where 
\begin{equation}
    \rho_{\rm min}=\left(1+f_s\right)\epsilon_{\rm ext} + g_{\rm outFOV}
\end{equation}
and 
\begin{equation}
\label{eq:eff_1}
    \rho_{max} =g_{\rm inFOV}+g_{\rm outFOV}
     + f_s\epsilon_{\rm glimpse},
\end{equation}
If the external triggering instrument is more or as sensitive as BAT, i.e. if $\epsilon_{\rm ext}\gtrsim1$, then $\rho_{\rm min}\gtrsim 1.56$. Using the values of $g_{\rm inFOV}$ and $g_{\rm outFOV}$ found before, and adopting $\epsilon_{\rm glimpse}=1$, $f_s=0.15$, we obtain $\rho_{max}\sim 4.86 $, meaning that we have a net maximum gain of $+386\%$ in the detection rate. The gain is $+345\%$ if we consider just sources inside the BAT FOV. The choice $\epsilon_{\rm glimpse}=1$ is conservative; indeed the mosaic technique can likely produce higher SNR than standard imaging performed in the same time interval, since the mosaic assigns to each sky image a relative weight that scales with the image SNR itself \citep{2025ApJ...988...23P}. 

\subsection{Gain in detection rate of arcminute-localized GRB by \texttt{BAT-GLIMPSE}+\texttt{NITRATES}}
The term $R_{\rm nitrates+glimpse}$ in \cref{eq:rho_eq} contains GRBs localized with arcminute precision, but also GRBs detected with \texttt{NITRATES} which are coarsely localized even if inside the BAT FOV. If we define $\rho_{\rm arcmin}$ the gain by \texttt{BAT-GLIMPSE}+\texttt{NITRATES} relative to arcmin-localized GRBs only, with respect to onboard BAT trigger, the following inequality is valid:
\begin{equation}
\label{eq:eff}
    \rho_{\rm arcmin}<g_{\rm inFOV}\lambda_{\rm arcmin}
     + f_s\epsilon_{\rm glimpse},
\end{equation}
where $\lambda_{\rm arcmin}$ is the fraction of all the GRBs occurring inside the BAT FOV detectable by \texttt{NITRATES} and for which an arcmin localization is attainable by \texttt{NITRATES} itself. Hence, the product $g_{\rm inFOV}\lambda_{\rm arcmin}$ represents the gain in detection rate of arcminute localized GRBs by \texttt{NITRATES}\footnote{under the condition that the external triggering instrument is not limiting the volume accessible by \texttt{NITRATES}.} with respect to the same rate obtained using standard imaging with \texttt{BAT-GLIMPSE}. This number can be obtained from \cref{fig:ts_vs_pc} counting the number of GRBs localized by \texttt{NITRATES} over the ones by \texttt{BAT-GLIMPSE}. However, this estimate is biased by the selection related to the sensitivity of the external triggering instrument. In the hypothesis that $g_{\rm inFOV}>\epsilon_{\rm ext}$, the external instrument limits the accessible volume selecting brighter and closer sources, therefore the ratio attainable from \cref{fig:ts_vs_pc} can be interpreted only as an upper limit for $g_{\rm inFOV}\lambda_{\rm arcmin}$. Thus we can arrive at the conservative estimate $g_{\rm inFOV}\lambda_{\rm arcmin}\lesssim 2.1$ and $\rho_{\rm arcmin}\lesssim2.25$, i.e. a net maximum gain of $+125\%$ in the detection rate of arcmin localized GRBs.

Similarly, a lower bound for  $\rho_{\rm arcmin}$ can be derived heuristically by computing the rate at which GCN circulars reporting arcminute-localized GRBs are published, obtained with imaging or mosaic, and compare it to the nominal \textit{Swift} GRB trigger rate. This fraction has been reported at the beginning of this section and corresponds to a gain of $\sim +17\%$. 
This represents a lower limit since: 1) the archival re-run of \texttt{BAT-GLIMPSE} has been done only on GRBs reported in GCN circulars and not systematically on the whole set of GRBs that triggered GUANO so far; 2) the detection horizon of the external triggering instrument may limit the volume detectable by \texttt{NITRATES}+\texttt{BAT-GLIMPSE}. Therefore, the actual value of $\rho_{\rm arcmin}$ is likely larger than $\sim17\%$. 

In conclusion, we arrive at a conservative range for the gain in detection rate of arcminute localized GRBs by \texttt{NITRATES}+\texttt{BAT-GLIMPSE} equal to $[+17\%, +125 \%]$ with respect to BAT triggering onboard, where we approach the upper bound in the optimistic scenario in which the triggering instrument has a detection horizon much larger than \texttt{NITRATES}. This is equivalent to saying that, in the hypothetical scenario where an offline search with \texttt{NITRATES} and \texttt{BAT-GLIMPSE} can be performed on the entirety of BAT TTE data, we would detect and localize to arcminute precision as much as double the amount of GRBs detected onboard by BAT.

\section{\texttt{BAT-GLIMPSE} and \texttt{ULTRA-Swift}: rapid response to early warning GW alerts} \label{ultra_swift}

\texttt{BAT-GLIMPSE} plays a critical role in high-energy transient searches and multi-messenger astronomy, as demonstrated by its deployment during the third part of the fourth
LVK observing run, providing a complementary coverage together with \texttt{NITRATES} in the monitoring of $\gamma-$ray counterparts of GWs, also during BAT slews.

Beginning in 2025, the LVK collaboration and the \textit{Swift} observatory initiated the \textit{ULTRA-Swift} project (Baylor et al., in prep.).
The project consists of creating a fast and direct communication channel between LVK and \textit{Swift} to stream in extreme low latency sky localization maps of GW signals detected before the merger itself. The system saves $\sim$20 s of latency by bypassing the orchestration performed by GWCelery \citep{2024PNAS..12116474C}, in charge of disseminating public notices and uploading GW candidates on GraceDB. Depending on the chirp mass of the system and the distance, at the design sensitivity of currently operating GW interferometers the alert can be streamed up to 30 - 60 s pre-merger for binary neutron stars \citep{2020ApJ...905L..25S,2020ApJ...902L..29N,2021ApJ...910L..21M,2022ApJ...927L...9K,2022PhRvD.106d2002B,2023ApJ...959...76C,2024arXiv240710263V} and 10 - 40 s for neutron star - black hole systems \citep{2022MNRAS.512.3878T}. As soon as the GW pipeline detects the source and produces the pre-merger sky localization map, the information is sent to the \textit{Swift} Mission Operation Center (MOC), where the optimal slew maneuver is calculated based on the algorithm presented in 
\cite{2024ApJ...975L..19T}. The optimization consists in maximizing the probability that the GW source is inside the BAT FOV at the GW merger time. As shown by \cite{2024ApJ...975L..19T}, this strategy implies that for the great majority of the cases \textit{Swift} would be still slewing at the GW merger time. This motivated the implementation of \texttt{BAT-GLIMPSE} for this project. 

After receiving the alert, the MOC takes $\mathcal{O}(1 \,\rm sec)$ of processing time to compute the slew, after which the slew command is uplinked to the spacecraft. After $\mathcal{O}(10 \,\rm sec)$ the spacecraft starts the slew maneuver. Concurrently, a GUANO command is also sent requesting 90 s of TTE BAT data around the GW trigger time, available for downlink at the next available ground station, along with a request to save a subset of data in the time interval $[t_{\rm merger},t_{\rm merger } +2\,s]$ that will immediately reach the ground through TDRSS communications. Approximately 1 hr is necessary to get the full 90 s dataset on the ground, allowing the data to be processed by the \textit{Swift} data center and made ready for being analyzed by \texttt{BAT-GLIMPSE}. The TDRSS system is in charge of transmitting to the \textit{Swift} MOC information about spacecraft telemetry. Therefore, the choice of the temporal window is limited by the risk of creating excessively large backlogs onboard of \textit{Swift}. The duration of 2 s is also motivated by the range of expected GW-GRB delay in the case of neutron star mergers \citep{delay}. However, pre-merger electromagnetic emission is also possible and detectable \citep{2012PhRvL.108a1102T,2021ApJ...921...92B,2023MNRAS.519.3923C,2026arXiv260214300S}.

All the early-warning events that triggered \textit{ULTRA-Swift} had large values of false alarm rates and none was confirmed by other online pipelines at merger time, indicating a likely non-astrophysical origin. No source has been detected by \texttt{BAT-GLIMPSE}. Nonetheless, these triggers were essential to test the coordination between the production of GW sky localization map from the LVK side and rapid response in the spacecraft automated commanding and data analysis from the \textit{Swift} side.

\section{Discussion and conclusions} \label{conclusions}

We have presented \texttt{BAT-GLIMPSE} \citep{batglimpse}, an open-source pipeline based on modules available in \texttt{BatAnalysis} \citep{batanalysis} for localizing transient $\gamma-$ray sources in \textit{Swift}-BAT time-tagged event data during both pointing observations and spacecraft slews. The pipeline combines automated identification of time seeds, coded-mask imaging, and mosaic techniques to recover the arcminute position of high-energy transients using GUANO data. The pipeline selects which method to use, among imaging and mosaic, based on the attitude information. The search for $\gamma-$ray transients is iterated over multiple trials, maximizing the detection probability. The automation introduced by \texttt{BAT-GLIMPSE} minimizes human intervention related to analysis configuration, making the pipeline usable at the latencies required for astrophysical follow-up.

We have analyzed 66 GRBs that span five years of \textit{Swift} operations, from the onset of GUANO operations (early 2020) through March 2026. \texttt{BAT-GLIMPSE} successfully recovers 43 events with positions consistent with published localizations. In particular, 88$\%$ of GRBs occurring during \textit{Swift} slews are recovered, demonstrating the effectiveness of the mosaic approach. We find that reliable imaging localizations are generally obtained for SNR $>$ 7. In $\sim$ 50$\%$ of the GRB triggers  for which \texttt{NITRATES} achieves arcminute-precision localization, \texttt{BAT-GLIMPSE} independently recovers the source position. Accounting for the sensitivity of \texttt{NITRATES} and the fraction of time spent by \textit{Swift} slewing, we estimate that \texttt{BAT-GLIMPSE}+\texttt{NITRATES} can increase the onboard trigger rate of GRBs localizable by BAT with arcminute precision by up to a factor of two. \texttt{BAT-GLIMPSE} not only supplements \texttt{NITRATES} during slews, where the more sensitive likelihood-based analysis cannot operate, but also provides a faster localization when the source is detectable in the imaging domain, reducing the the latency up to a factor 30-60. 

A key application of \texttt{BAT-GLIMPSE} is within the \textit{ULTRA-Swift} program (Baylor et al., in prep.), a low-latency connection between the LVK collaboration and the \textit{Swift} Mission Operations Center that enables spacecraft slews in response to pre-merger gravitational-wave alerts. Because the optimal observing strategy often places \textit{Swift} in the middle of a slew at merger time, \texttt{BAT-GLIMPSE} was integrated into the process to analyze BAT data collected during these intervals. During O4, several early-warning triggers activated the system. Although no astrophysical counterparts were detected, these tests successfully validated the end-to-end infrastructure and demonstrated the operational readiness of the system. Pre-merger sky localization is achievable  for the next decade of GW observations \citep{2022ApJ...935..139M}, and will be routinely available for the next generation GW instruments \citep{2021ApJ...917L..27N,2021PhRvD.104f4013T,2023ApJ...958L..43H,2023A&A...678A.126B,2024PhRvD.109d3021M}, like Einstein Telescope \citep{2010CQGra..27s4002P} and Cosmic Explorer \citep{2021arXiv210909882E}, and the implementation of \texttt{BAT-GLIMPSE} in cooperation with \textit{ULTRA-Swift} constitutes a proof-of-concept for rapid electromagnetic follow-up of pre-merger gravitational-wave alerts, directly informing the design of next-generation pipelines.

In conclusion, this work shows how the inclusion of \texttt{BAT-GLIMPSE} in the GUANO/\texttt{NITRATES} infrastructure has a significant impact in the context of time-domain and multi-messenger astronomy. It enables full exploitation of BAT as a survey, discovery and localizing instrument in response to externally triggered transients, including gamma-ray bursts, gravitational-wave events, high-energy neutrinos, fast X-ray transients, and fast radio bursts. The methodology presented here provides a directly transferable framework for coded-mask data analysis pipelines of current and forthcoming missions, like SVOM-ECLAIRs \citep{2014SPIE.9144E..24G} or THESEUS-XGIS \citep{2021SPIE11444E..2KL}, ensuring its relevance well beyond the \textit{Swift} era.

\begin{acknowledgments}
 S.R. acknowledges support from the Astrophysics Center for Multi-messenger Studies in Europe (ACME), funded under the European Union’s Horizon Europe Research and Innovation Program, Grant Agreement No. 101131928. The authors acknowledge the computational resources offered by the ROAR cluster (\url{https://icds.psu.edu/services/roar/roar-collab-cluster/}) of the PennState University, where both \texttt{NITRATES} and \texttt{BAT-GLIMPSE} run in parallel. This research has made use of data and/or software
provided by the High Energy Astrophysics Science Archive
Research Center (HEASARC), which is a service of the
Astrophysics Science Division at NASA/GSFC.
The authors thank Michael Moss for the useful comments.
\end{acknowledgments}

\software{Astropy \citep{astropy:2013, astropy:2018,astropy:2022}, Astropy/reproject \citep{astropy_reproject},  Astroquery \citep{astroquery}, {BatAnalysis \citep{batanalysis}}, {BAT-GLIMPSE} \citep{batglimpse},
          NumPy \citep{numpy}, Matplotlib \citep{matplotlib}, Scipy \citep{scipy}, HEASoft \citep{HEASoft}, swiftbat\_python, swifttools \citep{swifttools}, Histpy \citep{histpy_martinez_castellanos_2024_14262814}\\}

\startlongtable

\startlongtable
\begin{deluxetable*}{lrrrrrrrrrr}
\tabletypesize{\footnotesize}
\tablewidth{0pt}
\tablecolumns{11}
\tablecaption{Summary table of \texttt{BAT-GLIMPSE} results. The first column is the GCN ID reporting the GUANO localization. The symbol $(^{\dagger})$ indicates that the position of the GRB localized by GUANO  was later confirmed by the detection of the afterglow. In the square brackets we show the GCN that reports the confirmation of afterglow detection. The symbol $(^{\dagger\dagger})$ indicates that the position of the GRB localized by GUANO is consistent with IPN triangulation. In bold we indicate those GCN circulars produced after \texttt{BAT-GLIMPSE} started to work in real-time.  RA and Dec are the coordinates reported in the GUANO GCN. PC indicates the partial coding fraction of the GRB at the trigger time. A \checkmark symbol in mosaic column means the source if found with mosaic. A \checkmark in the slew column means that the GCN mentions that the detection was during a slew of \textit{Swift}. $\sqrt{TS}$ is coming from \texttt{NITRATES} analysis. The SNR column reports the SNR indicated in the GCN (if any) and the SNR obtained with \texttt{BAT-GLIMPSE}. If the SNR \texttt{BAT-GLIMPSE} is empty, no source is found. If the slew is marked with a \checkmark, but mosaic is not, it means that the imaging found the source on a time window $<0.2$ s. The cases marked with [*] are the only two ones during slew not found by mosaic. All the cases not found by imaging either have an SNR$<7$ reported in the GCN, or the position is only reported by \texttt{NITRATES}. The last column reports the angular separation between the source found by \texttt{BAT-GLIMPSE} and the one reported in the GUANO GCN. GCN 29947 has a second GCN 29969 reporting that the IPN localizes the source outside the BAT FOV, so likely the source reported by GUANO was an artifact. In the case of GCN 43114, the first row is relative to the analysis done for the Fermi trigger time, and the second one for the later CALET trigger on the same GRB. \label{tab:gua--}}
\tablehead{
\colhead{GCN} & \colhead{RA} & \colhead{Dec} & \colhead{Trigger Time} & \colhead{PC} & \colhead{$\sqrt{TS}$} & \multicolumn{2}{c}{SNR} & \colhead{Mosaic} & \colhead{Slew} & \colhead{Separation} \\
\cline{7-8}
\colhead{} & \colhead{deg} & \colhead{deg} & \colhead{UTC} & \colhead{} & \colhead{} & \colhead{GCN} & \colhead{\texttt{BAT-GLIMPSE}} & \colhead{} & \colhead{} & \colhead{arcmin}
}
\startdata
    \href{https://gcn.nasa.gov/circulars/27107}{27107}  & 311.43780 & -11.65800 &    2020-02-16T09:07:25.030 &            0.39 &                   -- &        7.5 &           6.90 &     -- &   -- &            1.09 \\
  \href{https://gcn.nasa.gov/circulars/27259}{27259}  & 333.89280 & -42.94430 &    2020-02-28T06:58:33.810 &            0.39 &                   -- &       40.0 &          44.73 &     -- & \checkmark &            0.50 \\
  \href{https://gcn.nasa.gov/circulars/2744}{27444}$^{\dagger\dagger}$  [27449] &  31.72030 & -31.81600 &    2020-03-25T03:18:31.000 &            0.12 &                   -- &        7.5 &           6.72 &     -- &   -- &            0.58 \\
  \href{https://gcn.nasa.gov/circulars/27497}{27497}$^{\dagger\dagger}$  [27505]&  62.78940 & -51.53260 &    2020-04-05T03:53:38.000 &            0.13 &                12.40 &        4.0 &           5.63 &     -- &   -- &         $>60$ \\
  \href{https://gcn.nasa.gov/circulars/28013}{28013} & 242.09520 &  53.46780 &    2020-06-23T03:18:00.000 &            0.63 &                 9.80 &        6.4 &             -- &     -- &   -- &              -- \\
  \href{https://gcn.nasa.gov/circulars/28103}{28103} $^{\dagger}$ [28150]& 196.86560 & -51.64030 &    2020-07-14T18:35:05.000 &            0.26 &                   -- &       11.5 &          31.49 &     -- &   -- &            2.65 \\
  \href{https://gcn.nasa.gov/circulars/28237}{28237} $^{\dagger}$ [28241] &  15.94060 & -73.84590 &    2020-08-09T15:41:27.000 &            0.34 &                   -- &       10.4 &          10.76 &     -- &   -- &            1.08 \\
  \href{https://gcn.nasa.gov/circulars/28583}{28583} $^{\dagger}$ [28648]& 161.74400 &  46.10100 &    2020-10-08T10:37:36.000 &            0.03 &                13.22 &        4.2 &           5.53 &     -- &   -- &         $>60$ \\
  \href{https://gcn.nasa.gov/circulars/28971}{28971}  & 339.35400 & -49.24600 &    2020-11-28T17:54:11.000 &            0.76 &                   -- &         -- &           6.35 &     -- & \checkmark &            6.29 \\
   \href{https://gcn.nasa.gov/circulars/29059}{29059}  & 201.48720 &  36.31160 &    2020-12-16T19:00:31.240 &            0.50 &                   -- &        8.9 &          11.75 &     -- &   -- &            2.03 \\
  \href{https://gcn.nasa.gov/circulars/29701}{29701}  & 259.66400 &  15.67700 &    2021-03-23T12:03:33.090 &            0.13 &                11.77 &         -- &             -- &     -- &   -- &              -- \\
  \href{https://gcn.nasa.gov/circulars/29877}{29877}  & 270.81700 &  56.82800 &    2021-04-21T10:54:44.810 &            0.59 &                   -- &        9.7 &           9.45 & \checkmark & \checkmark &            2.46 \\
  \href{https://gcn.nasa.gov/circulars/29947}{29947} & 132.85300 &   4.58200 &    2021-05-06T00:39:48.540 &            0.84 &                   -- &        6.8 &             -- &     -- &   -- &              -- \\
  \href{https://gcn.nasa.gov/circulars/29984}{29984}$^{\dagger\dagger}$ [28121]  & 139.34800 & -16.71200 &    2020-07-16T01:26:37.760 &            0.02 &               110.96 &        8.5 &           8.02 &     -- &   -- &            0.83 \\
  \href{https://gcn.nasa.gov/circulars/29985}{29985}  &  35.59400 &  56.01500 &    2020-12-28T15:21:45.000 &            0.56 &                   -- &       13.1 &          14.81 &     -- &   -- &            0.62 \\
  \href{https://gcn.nasa.gov/circulars/30130}{30130}$^{\dagger}$ [30234]  &  15.73200 &  -6.46700 &    2021-06-05T14:55:58.000 &            0.53 &                   -- &        8.4 &           8.71 &     -- & \checkmark &            1.27 \\
  \href{https://gcn.nasa.gov/circulars/30134}{30134}$^{\dagger}$ [30172]& 170.90400 &   0.71800 &    2021-06-06T03:56:02.000 &            0.15 &                15.00 &        5.2 &             -- &     -- &   -- &              -- \\
  \href{https://gcn.nasa.gov/circulars/30302}{30302} & 233.11700 & -26.21300 &    2021-06-22T01:32:35.900 &            0.35 &                   -- &        8.1 &           8.86 &     -- &   -- &            3.94 \\
  \href{https://gcn.nasa.gov/circulars/30325}{30325}$^{\dagger}$ [30341] & 221.61990 &  -1.15120 &    2021-06-26T08:16:15.900 &            0.63 &                   -- &        7.6 &          23.83 & \checkmark & \checkmark &            4.58 \\
  \href{https://gcn.nasa.gov/circulars/30393}{30393}$^{\dagger}$ [30431]& 312.01240 &  13.30790 &    2021-07-06T08:17:49.900 &            0.82 &                   -- &       19.0 &          22.77 & \checkmark & \checkmark &            1.97 \\
  \href{https://gcn.nasa.gov/circulars/30732}{30732}$^{\dagger}$ [30745]& 174.91810 &  55.78580 &    2021-08-27T09:59:24.340 &            0.53 &                   -- &        8.3 &           8.48 &     -- &   -- &            0.96 \\
  \href{https://gcn.nasa.gov/circulars/31006}{31006} &  28.05300 &  -6.97300 &    2021-10-24T01:34:11.400 &            0.10 &                10.40 &         -- &           5.74 &     -- &   -- &         $>60$ \\
  \href{https://gcn.nasa.gov/circulars/31049}{31049}$^{\dagger}$ [31068]& 343.64300 & -53.23600 &    2021-11-06T04:37:31.200 &            0.05 &                19.10 &         -- &           5.68 &     -- &   -- &          $>60$ \\
  \href{https://gcn.nasa.gov/circulars/31334}{31334} & 295.04400 &  23.13300 &    2021-12-29T03:29:15.000 &            0.75 &                50.00 &         -- &          19.40 &     -- &   -- &            1.60 \\
  \href{https://gcn.nasa.gov/circulars/31402}{31402}$^{\dagger}$  [31410]& 169.76100 &  34.13900 &    2022-01-07T14:45:31.000 &            0.08 &                   -- &        9.0 &             [*] &     -- & \checkmark &              -- \\
  \href{https://gcn.nasa.gov/circulars/31501}{31501}  & 286.45700 &  16.79100 &    2022-01-17T04:45:55.000 &            0.07 &                   -- &        6.1 &           6.91 &     -- &   -- &            1.00 \\
  \href{https://gcn.nasa.gov/circulars/31572}{31572} & 345.77800 &  61.82600 &    2022-02-10T23:56:38.000 &              -- &                   -- &       11.4 &             \multicolumn{4}{c}{corrupted event file}\\
  \href{https://gcn.nasa.gov/circulars/31746}{31746}  & 290.06900 &  40.25300 &    2022-03-10T22:23:51.000 &            0.17 &                 9.70 &         -- &           6.16 &     -- &   -- &         $>60$ \\
  \href{https://gcn.nasa.gov/circulars/31919}{31919}  & 224.32900 & -17.51400 &    2022-04-18T17:16:21.000 &            0.27 &                10.40 &         -- &           6.56 &     -- &   -- &        $>60$ \\
  \href{https://gcn.nasa.gov/circulars/32023}{32023}$^{\dagger}$ [32052]& 287.44600 &  17.73900 &    2022-05-11T13:41:56.000 &            0.07 &                   -- &       10.0 &          11.57 &     -- & \checkmark &            2.59 \\
  \href{https://gcn.nasa.gov/circulars/32167}{32167}  &  28.47400 & -54.74100 &    2022-06-06T01:03:28.000 &            0.43 &                10.70 &         -- &             -- &     -- &   -- &              -- \\
  \href{https://gcn.nasa.gov/circulars/32345}{32345}$^{\dagger}$ [32350]& 248.65200 &  36.33800 &    2022-07-08T02:06:55.110 &            0.03 &                41.40 &         -- &           5.99 &     -- &   -- &            2.74 \\
  \href{https://gcn.nasa.gov/circulars/32375}{32375}$^{\dagger}$ [32385] & 221.59200 &  20.65700 &    2022-07-10T03:29:34.000 &            0.75 &                   -- &       20.0 &          21.04 & \checkmark & \checkmark &            0.90 \\
    \href{https://gcn.nasa.gov/circulars/32506}{32506}$^{\dagger}$ [32513]&  24.25500 & -41.58180 &        2022-08-31T13:56:33.000 &              -- &        -- &       15.0 &        \multicolumn{4}{c}{no guano data} \\
  \href{https://gcn.nasa.gov/circulars/32543}{32543} & 216.04480 &  -8.85510 &    2022-09-09T06:52:13.000 &            0.23 &                 8.90 &         -- &             -- &     -- &   -- &              -- \\
  \href{https://gcn.nasa.gov/circulars/32941}{32941}  &  54.78800 & -16.70100 &    2022-11-15T09:46:15.000 &            0.07 &                27.70 &         -- &           5.94 &     -- &   -- &         $>60$ \\
  \href{https://gcn.nasa.gov/circulars/33027}{33027}  & 251.15500 &  42.67100 &    2022-12-06T12:22:47.000 &            0.08 &                   -- &        6.6 &             [*] &     -- & \checkmark &              -- \\
  \href{https://gcn.nasa.gov/circulars/33062}{33062}  & 155.05600 &  46.87300 &    2022-12-15T04:14:27.000 &            0.47 &                11.80 &        5.5 &           6.05 &     -- &   -- &         $>60$ \\
  \href{https://gcn.nasa.gov/circulars/33132}{33132}$^{\dagger}$ [33139]& 336.26000 &  25.13800 &    2022-12-31T21:46:05.000 &            0.22 &                14.00 &         -- &           6.18 &     -- &   -- &            1.87 \\
  \href{https://gcn.nasa.gov/circulars/33672}{33672} &  41.70400 &   1.11900 &    2023-04-21T09:42:46.370 &            0.45 &                   -- &       10.5 &          10.48 & \checkmark & \checkmark &            4.32 \\
  \href{https://gcn.nasa.gov/circulars/34159}{34159}  & 173.19920 & -16.41210 &    2023-07-07T03:22:51.000 &            0.61 &                   -- &       10.0 &          23.31 &     -- &   -- &            0.21 \\
  \href{https://gcn.nasa.gov/circulars/34683}{34683}  & 267.34900 &  74.48400 &    2023-09-13T07:56:12.160 &            0.13 &                19.70 &         -- &           6.90 &     -- &   -- &            1.06 \\
  \href{https://gcn.nasa.gov/circulars/34824}{34824}  &  86.19100 &  56.66600 &    2023-10-17T08:05:03.300 &            0.03 &                15.00 &         -- &           6.63 &     -- &   -- &         $>60$ \\
  \href{https://gcn.nasa.gov/circulars/35022}{35022}$^{\dagger}$ (SGR)& 238.81800 & -54.07700 &    2023-11-11T21:35:11.400 &            0.84 &                29.20 &         -- &             -- &     -- &   -- &              -- \\
  \href{https://gcn.nasa.gov/circulars/35409}{35409}  &  27.32100 &  77.89700 &    2023-12-20T16:10:17.400 &            0.29 &                28.32 &         -- &           9.43 &     -- &   -- &            0.72 \\
  \href{https://gcn.nasa.gov/circulars/35733}{35733}$^{\dagger}$ [35852]& 132.23700 &  -6.40580 &    2024-02-15T15:33:00.410 &            0.84 &                   -- &       13.9 &          13.64 &     -- &   -- &            0.46 \\
  \href{https://gcn.nasa.gov/circulars/35880}{35880}  &   6.27800 & -43.90400 &    2024-03-05T09:49:21.260 &            0.78 &                   -- &        9.7 &          11.01 & \checkmark & \checkmark &            1.80 \\
  \href{https://gcn.nasa.gov/circulars/35948}{35948} & 272.18000 &  39.18200 &    2024-03-17T10:54:24.000 &            0.84 &                49.27 &         -- &          23.94 &     -- &   -- &            8.11 \\
  \href{https://gcn.nasa.gov/circulars/36125}{36125}  & 132.31700 & -27.07000 &    2024-04-15T03:10:49.351 &            0.11 &                16.20 &         -- &             -- &     -- &   -- &              -- \\
  \href{https://gcn.nasa.gov/circulars/36532}{36532}  & 355.68790 & -32.72150 &    2024-05-21T15:08:57.060 &            0.84 &                74.60 &         -- &          32.08 &     -- &   -- &            0.38 \\
  \href{https://gcn.nasa.gov/circulars/36628}{36628}$^{\dagger}$ [36631]&  40.23400 & -60.63700 &    2024-06-06T13:33:06.500 &            0.04 &                28.90 &         -- &           6.99 &     -- &   -- &         $>60$ \\
  \href{https://gcn.nasa.gov/circulars/36672}{36672}$^{\dagger}$ [36683]& 326.14130 &  38.59480 &    2024-06-15T17:51:45.000 &            0.54 &                   -- &       12.0 &           9.53 &     -- & \checkmark &            6.48 \\
  \href{https://gcn.nasa.gov/circulars/36945}{36945}  &  97.78600 & -37.18000 &    2024-07-27T06:00:08.960 &            0.15 &                45.60 &         -- &           7.33 &     -- &   -- &            0.47 \\
  \href{https://gcn.nasa.gov/circulars/38091}{38091}$^{\dagger}$ [38098]&  66.23810 & -49.75470 &    2024-11-05T16:06:04.000 &            0.22 &                   -- &       11.0 &          15.42 &     -- &   -- &            0.16 \\
  \href{https://gcn.nasa.gov/circulars/38840}{38840}  & 302.52000 &  45.56400 &    2025-01-07T05:02:30.290 &            0.01 &                21.80 &         -- &           6.11 &     -- &   -- &         $>60$ \\
  \href{https://gcn.nasa.gov/circulars/39744}{39744}$^{\dagger}$ [39758] & 219.63000 &  14.88600 &    2025-03-16T08:59:57.690 &            0.08 &                17.70 &         -- &           7.64 &     -- &   -- &            0.85 \\
  \href{https://gcn.nasa.gov/circulars/39979}{\textbf{39979}}$^{\dagger}$ [40079]&  48.52543 &  62.30596 &    2025-03-31T05:49:17.010 &            0.77 &                   -- &       11.6 &          14.56 & \checkmark & \checkmark &            5.20 \\
  \href{https://gcn.nasa.gov/circulars/40563}{40563}  & 319.63300 &  52.73700 &    2025-05-27T13:58:32.410 &            0.30 &                11.60 &         -- &             -- &     -- &   -- &              -- \\
  \href{https://gcn.nasa.gov/circulars/40600}{40600}$^{\dagger}$ [40606]&  90.07090 & -57.28920 &    2025-06-01T17:29:35.540 &            0.31 &                   -- &        8.7 &          10.04 &     -- &   -- &            0.75 \\
  \href{https://gcn.nasa.gov/circulars/41110}{\textbf{41110}} &  44.11790 &  26.37890 &    2025-07-17T13:44:22.000 &            0.45 &                   53.47 &       26.3 &          26.37 &     -- & --  &            0.00 \\
  \href{https://gcn.nasa.gov/circulars/41185}{\textbf{41185}}  & 260.94500 &  16.20490 &    2025-07-28T09:25:45.570 &            0.66 &                   16.4 &        8.1 &           8.23 &     -- & --  &            0.03 \\
  \href{https://gcn.nasa.gov/circulars/42244}{42244}  &  46.90100 &   8.80000 &    2025-10-13T18:17:00.390 &            0.24 &                27.60 &         -- &          11.77 &     -- &   -- &            1.10 \\
  \href{https://gcn.nasa.gov/circulars/42245}{\textbf{42245}}  &   4.78780 &  45.18940 &    2025-10-11T03:41:40.090 &            0.73 &                   -- &       23.5 &          22.67 &     -- & \checkmark &            0.00 \\
  \href{https://gcn.nasa.gov/circulars/42263}{\textbf{42263}}  &  14.59540 & -35.58020 &    2025-10-14T04:47:21.170 &            0.72 &                   19.78 &        8.5 &           8.48 &     -- & -- &            0.85 \\
  \href{https://gcn.nasa.gov/circulars/42568}{42568}$^{\dagger}$ [42551] & 146.82100 &  16.10400 &    2025-11-03T04:46:28.110 &            0.02 &                61.60 &         -- &             -- &     -- &   -- &              -- \\
  \href{https://gcn.nasa.gov/circulars/43033}{43033}$^{\dagger}$ [43050]& 228.45300 &  29.04400 &    2025-12-08T10:17:17.290 &            0.33 &                17.60 &         -- &           7.53 &     -- &   -- &            1.98 \\
  \href{https://gcn.nasa.gov/circulars/43114}{\textbf{43114}}$^{\dagger}$ [43155]& 122.65260 & -36.73950 &    2025-12-15T04:23:36.280 &            0.16 &                  14.6 &        6.5  &           7.76 &     -- & -- &            0.58 \\
  &  &  &  & -- &  --  &   27.1 &           7.9 &     \checkmark & \checkmark &            4.25 \\
  \href{https://gcn.nasa.gov/circulars/43377}{\textbf{43377}}$^{\dagger}$ [43426]& 229.57340 &  51.97470 &    2026-01-11T16:33:14.010 &            0.63 &                   36.98 &       19.7 &          23.13 &     -- & -- &            0.48   \\
\enddata
\end{deluxetable*}


\bibliography{sample702}{}
\bibliographystyle{aasjournal}

\end{document}